\documentclass[aps,groupedaddress,pra,twocolumn,amsmath,amssymb,superscriptaddress]{revtex4-1}

\usepackage[english]{babel} 
\usepackage[utf8]{inputenc}
\usepackage[T1]{fontenc}

\usepackage[normalem]{ulem}

\usepackage{bbm}
\usepackage{verbatim}
\usepackage{graphicx}
\usepackage{times}
\usepackage{epsfig}
\usepackage{graphicx}
\usepackage{bm}
\usepackage{txfonts}
\usepackage{dsfont}
\usepackage{color}
\usepackage{braket}

\def\tf{t_{\rm f}}

\usepackage{hyperref}
\hypersetup{
colorlinks=true,
linkcolor=blue,          % color of internal links
citecolor=blue,          % color of links to bibliography
filecolor=magenta,       % color of file links
urlcolor=blue
}

\begin{document}

\selectlanguage{english}

\title{ Quantum state preparation for coupled period tripling oscillators}

\author{Niels L\"orch}
\affiliation{Department of Physics, University of Basel, Klingelbergstrasse 82, CH-4056 Basel, Switzerland}
\author{Yaxing Zhang}
\affiliation{Department of Physics, Yale University, New Haven, Connecticut 06511, USA}
\author{Christoph Bruder}
\affiliation{Department of Physics, University of Basel, Klingelbergstrasse 82, CH-4056 Basel, Switzerland}
\author{M. I. Dykman}
\affiliation{Department of Physics and Astronomy, Michigan State University, East Lansing, Michigan 48824, USA}

\date{\today}

\begin{abstract}
We investigate the quantum transition to a correlated state of coupled
oscillators in the regime where they display period tripling in
response to a drive at triple the eigenfrequency.
Correlations are formed between the discrete oscillation phases of individual oscillators. 
The evolution toward the ordered state is accompanied by the transient breaking of the symmetry between seemingly  equivalent configurations. 
We attribute this to the
nontrivial geometric phase that characterizes period tripling.  We
also show that the Wigner distribution of a single damped quantum
oscillator can display a minimum at the classically stable
zero-amplitude state. 

\end{abstract}

\maketitle

\paragraph*{Introduction.--}

The adiabatic theorem in quantum mechanics~\cite{born1928} states that
a quantum system in the instantaneous ground state of a
time-dependent Hamiltonian will approximately remain there
if the Hamiltonian changes slowly compared to the gap to the first excited state.
Recently the
adiabatic dynamics in many-body systems has been extensively
studied with arrays of qubits~\cite{Lloyd,Boixo2014,Boixo2016}.
One promising application is adiabatic quantum
computing, where the initial Hamiltonian is well-understood, so
that initialization of its ground state is straightforward, and the
final Hamiltonian encodes the cost function of an optimization
problem that is hard to solve on a classical
computer~\cite{Kadowaki1998,Farhi2001,Childs2001}.

The interest in adiabatic many-body dynamics has now extended to
systems of quantum oscillators~\cite{Goto2016,Nigg2017,Puri2017,
  savona2017,Mamaev2018,Goto2018,Zhao2018,Dykman2018,goto2018q,rota2019}. This
was triggered by the observation how, with turning on parametric
driving close to twice the oscillator eigenfrequency, the ground state
of a single oscillator adiabatically connects to the cat
state~\cite{Goto2016,Zhang2017,Puri2017npj, Goto2018,wang2019}, and
how this can be used for adiabatic quantum computing with oscillator
arrays~\cite{Fitzpatrick2016,Ma2019}.  Coupled coherent parametrically
driven oscillators can go through a quantum phase transition into a
correlated state (a ``time-crystal'' effect with no disorder)
~\cite{Dykman2018}. 

Parametric oscillators
can be mapped ~\cite{Goto2016} onto an ``Ising
machine'', which has recently been demonstrated in the classical regime with 100-2000 optical spins
 \cite{McMahon2016, Inagaki2016}.

\begin{figure}[t!]
\includegraphics[width=0.18\textwidth]{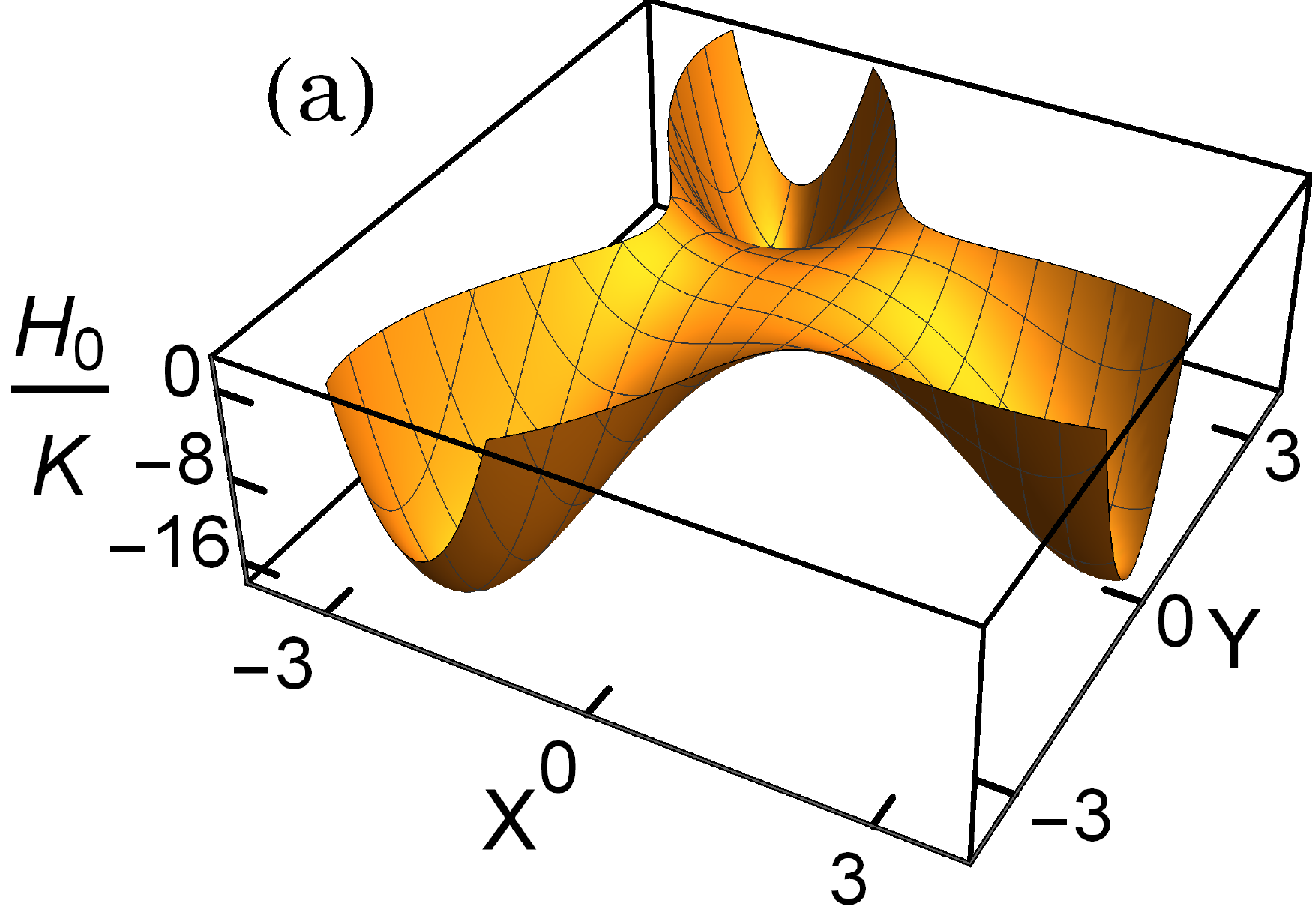}\hfill
\includegraphics[width=0.16\textwidth]{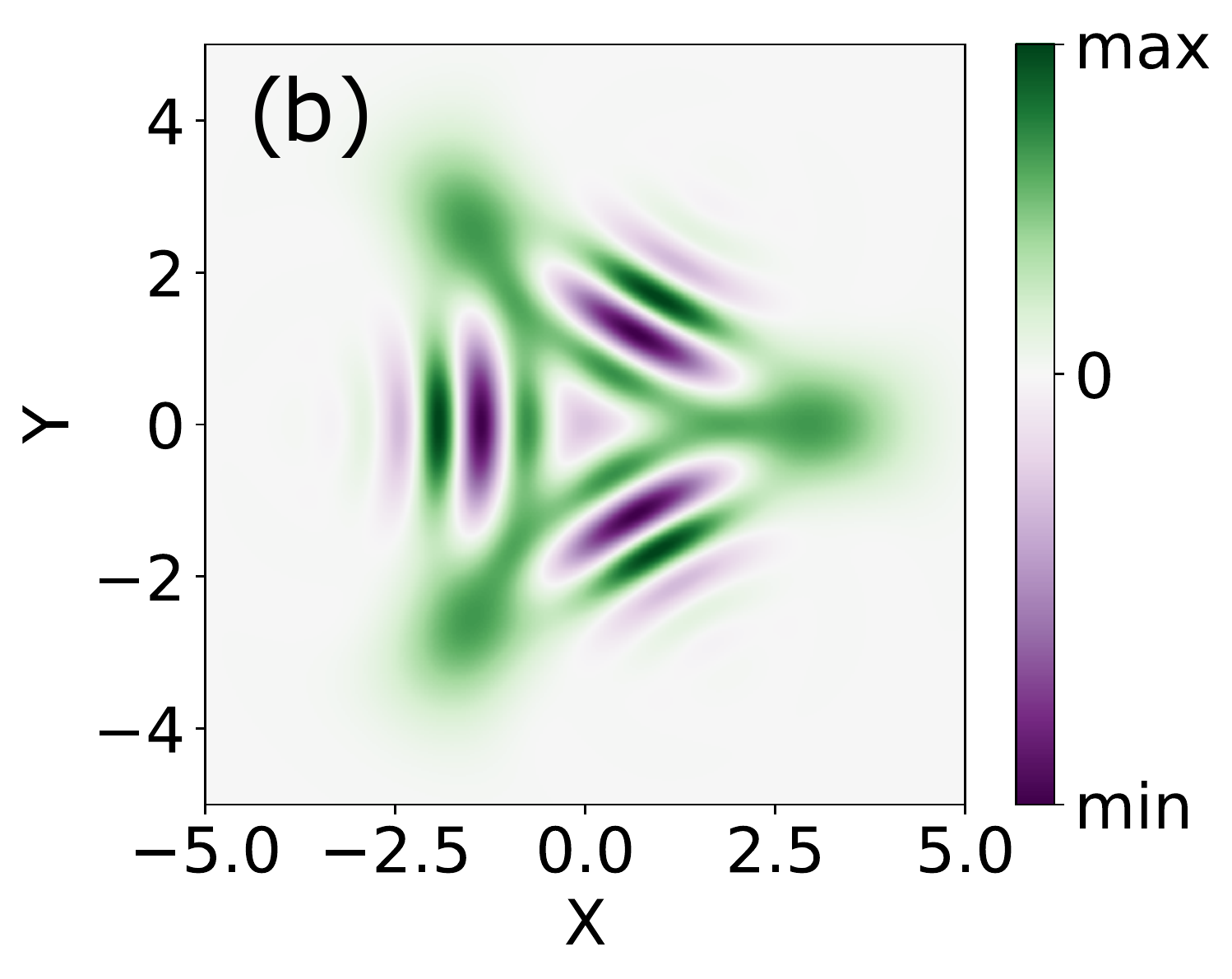}\hfill
\includegraphics[width=0.14\textwidth]{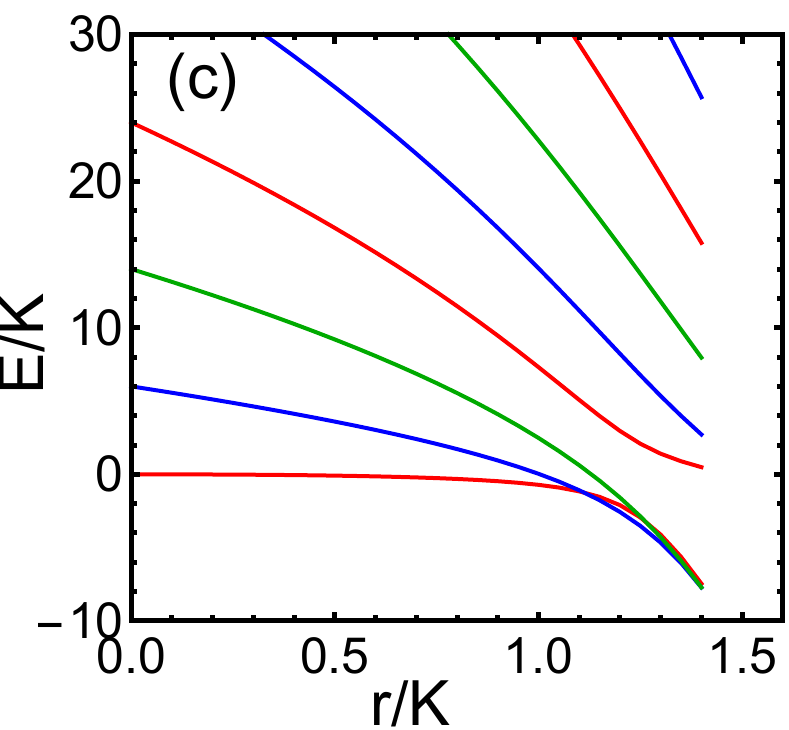} 
\caption{(a) Classical phase-space energy surface of a single
  oscillator in the rotating frame for driving at triple the
  eigenfrequency. The plot corresponds to $H_0$, Eq.~(\ref{eq:H_0}),
  in units of the Kerr parameter $K$; $X$ and $Y$ are the scaled
  coordinate and momentum, $r=1.4K$, and $\Delta=0$.  In the main
  text, the minima are enumerated counter-clockwise as $j=0,1,2$
  starting with the minimum on the axis $Y=0$. (b) Wigner distribution
  in the lowest fully symmetric eigenstate of $H_0$ for $r=1.4K$ and
  $\Delta=0$. (c) Eigenvalues of $H_0$ as functions of $r/K$ where
  $\Delta=\Delta_{\rm ini}(1-r/r_{\max})$, $\Delta_{\rm ini}=6K$. For
  $r=0$ the spectrum is that of a weakly anharmonic oscillator and the
  levels are color-coded as $n=3k$ (red), $3k+1$ (blue), $3k+2$
  (green), with $k=0,1,2,...$. With the increasing $r/K$ the levels
  with different $k$ merge into triples of tunnel-split intrawell
  levels of $H_0$.}
\label{fig:cat}
\end{figure}

The many-body dynamics of driven coupled oscillators can be radically
different if the driving frequency is close to triple the oscillator
eigenfrequency. An isolated oscillator can display period tripling in
this case. A particular feature of the effect is the geometric phase
\cite{Zhang2017} between the quantum states at the minima of the
effective oscillator Hamiltonian in Fig.~\ref{fig:cat} first noticed
in Ref.~\onlinecite{Guo2013a}. It can be thought of as resulting from
a ``magnetic field'' that pierces the oscillator phase space.

In this paper we study how the geometric phase of the quantum period
tripling and the high degeneracy of the period-3 states affect the
dynamics of coupled quantum oscillators. Specifically, we study how
the system goes into a coherent many-body state as the driving field
is slowly turned on and tuned close to resonance. The results refer to
a one-dimensional oscillator array with either attractive or repulsive
couplings. Such couplings favor, respectively, the same or different
phases of the period-3 oscillations and are analogous to ferro- or
antiferromagnetic coupling in the case of spins. The case of
antiferromagnetic coupling is particularly interesting because
multiple configurations can lead to neighboring oscillators having
different phases. We note that, because of the geometric phase, the
oscillator chain cannot be simply mapped on a chain of spin-1
particles.

We also study the stationary distribution of a single weakly-damped
oscillator in the ultra-quantum regime to explore whether period
tripling can qualitatively change this distribution compared to what
would be expected in the semiclassical limit. The very possibility of
such a change is a consequence of the peculiar semiclassical dynamics
where the unstable period-3 states approach the stable state with the
increasing drive, but do not merge with this state.

\paragraph*{Physical setup and Hamiltonian.--}

We study arrays of $N$ coupled driven oscillators. The Hamiltonian
\begin{align}
\label{eq:full_Hamiltonian}
H=H_s+H_d+H_i
\end{align}
consists of the Hamiltonian of the undriven oscillators ($H_s$), the
driving term ($H_d$), and the interaction
($H_i$). We assume that all oscillators are identical and have
inversion symmetry, and we keep in $H_s$ the lowest-order intrinsic
nonlinearity (called Duffing or Kerr nonlinearity).  In the frame that
rotates at 1/3 the drive frequency $\omega_F$ and in the familiar
rotating wave approximation (RWA)~\cite{Walls2008}
\begin{align}
\label{eq:Hamiltonian_chain_uncoupled}
H_s= \sum_n \Delta a_n^\dagger a_n + K (a_n^\dagger)^2 a_n^2\:,
\end{align}
where $a_n$ and $a_n^\dagger$ are the ladder operators of the $n$th oscillator. In Eq.~(\ref{eq:Hamiltonian_chain_uncoupled}) we introduced the detuning $\Delta=\omega_0-\omega_F/3$ of the drive with respect
to the oscillator eigenfrequency $\omega_0$; $K$ is the nonlinearity
parameter, and we
set $\hbar = 1$.  

The Hamiltonian that describes the driving
\begin{align}
\label{eq:driving}
H_d =  -r \sum_n  \left[a_n^3 + (a_n^\dagger)^3 \right]
\end{align}
corresponds to the energy of an oscillator in the driving field, which
is proportional to the field multiplying the cube of the oscillator
coordinate, with $r$ being the scaled field amplitude. The term
(\ref{eq:driving}) can arise also from a coupling linear in the
coordinate or momentum taking into account the oscillator
nonlinearity, cf. \cite{Zhang2017}.

From Eqs.~(\ref{eq:Hamiltonian_chain_uncoupled}) and
(\ref{eq:driving}), we can write the RWA Hamiltonian $H_0=H_s+H_d$ of
an individual oscillator as
\begin{align}
\label{eq:H_0}
H_0 = &\frac{1}{2}\Delta (X^2+Y^2-1) +\frac{1}{4}K[(X^2 + Y^2-2)^2 -1]\nonumber\\
&- r(X^3-3YXY)/\sqrt 2\:,
\end{align}
where $X$ and $Y$ correspond to the scaled coordinate and momentum,
$X=(a^\dagger+a)/\sqrt 2$ and $Y=i(a^\dagger-a)/\sqrt 2$.

The classical phase-space energy surface corresponding to $H_0$ is
shown in Fig.~\ref{fig:cat} along with an example of the Wigner
distribution. The Hamiltonian has a three-fold symmetry in the
oscillator phase space, a feature of period tripling. The three minima
away from $X=Y=0$ emerge for $r^2 > 8K(\Delta-2K)/9$. Classically, they
correspond to different phases $\theta = 0,\,2\pi/3$, and $4\pi/3$ of
the period-3 oscillations.

We assume linear coupling between the oscillators. After an RWA it is described by the interaction Hamiltonian
\begin{align}
\label{eq:coupling}
H_i=-\sum_{m\neq n}^{N} V_{mn} a_m^\dagger a_n\: .
\end{align}
To reveal the novel features of the many-body dynamics coming from
period tripling, we consider the simplest model of the oscillator
array: a nearest-neighbor coupling, $V_{mn}=V\delta_{m,n\pm 1}$, and
periodic boundary conditions. For the ``ferromagnetic'' and
``antiferromagnetic'' cases, $V>0$ and $V<0$, respectively. Below we
loosely use the term ``energy'' for the eigenvalues of the Hamiltonian
$H$.

The Hamiltonian is invariant under simultaneous rotation of all
oscillators by $-2\pi/3$, which is realized by the unitary operator
${\mathbb N}_3 = \exp\left[-(2\pi i/3) \sum\nolimits_n a_n^\dagger
  a_n\right]$.  The other symmetry operations are translation
${\mathbb T}^\dagger a_n {\mathbb T} = a_{n+1}$ and reversing
${\mathbb R}^\dagger a_n {\mathbb R} = a_{N+1-n}$ the order of the
oscillators.

\paragraph*{Measurement of states.--}
For the classification and measurement of the states we use
the resolution of unity with coherent states, $ \hat I = \frac 1 \pi
\int_0^\infty |\alpha| \,\mathrm{d} |\alpha| \int_0^{2 \pi} \mathrm{d} \theta  \ket \alpha
\bra \alpha$, with $\alpha = |\alpha|\exp(i\theta)$, 
to define the measurement operators
\begin{align}
\label{eq:E_theta}
E(\theta) &= \frac 1 \pi \int_0^\infty |\alpha|\, \mathrm d |\alpha| \int_{-\theta}^{\theta}
   \mathrm{d} \theta\ket {\alpha} \bra {\alpha},
\end{align} 
In terms of the oscillator Fock states $\ket{k}$ in the absence of driving, $E(\theta) = \frac 1 \pi  \sum_{k,k'=0}^\infty \frac{ \Gamma \bigl( (k+k'+2)/2 \bigr) }
 {\sqrt{k!k'!}} \cdot
\frac{\sin \left[ (k-k') \theta \right] }{k-k'} \cdot \ket k \bra k'$, where
$\Gamma(x)$ is the Gamma-function, and we use the convention
$\theta >0$.  The approximate effect of $E(\theta)$ is a projection on
the sector of phase space bounded by the polar angles $-\theta$ and
$\theta$.  As the coherent states do not form an orthogonal basis,
$E(\theta)$ is not a projector, but corresponds to a more general form
of measurement that can be described in the framework of Positive
Operator Valued Probability Measures
(POVMs)~\cite{Busch1996}.  

We define $P_0 = E(\pi/3)$, corresponding to the third of phase space
limited by the polar angles $\pm \pi/3$. For one oscillator, where the
phase-space rotation operator is $N_3=\exp[-(2\pi i/3) a^\dagger a]$,
we define the rotated operators $P_1= N_3^\dagger P_0 N_3$ and
$P_2= N_3^\dagger P_1 N_3$ corresponding to, respectively, the sectors
rotated by $2 \pi /3$ and $4 \pi /3$.  As $P_0+P_1+P_2 =\hat I$, the
$P$-operators form a POVM, and we define the corresponding
probabilities as $p_j=\bra \psi P_j \ket \psi$, where $\ket\psi$ is
the oscillator wave function. These definitions naturally generalize
to arrays of oscillators. For two oscillators the probability of the
first oscillator to be in sector $j$ and the second oscillator to be
in sector $k$ is $p_{jk}=\bra \psi P_j \otimes P_k \ket \psi$. In the
general case $p_{j_1...j_N}=\bra \psi \Pi_{n=1}^N P_{j_n} \ket \psi$.

\paragraph*{Quasi-adiabatic state preparation.--}
We will assume that each oscillator is initialized in the vacuum state, $\ket \psi_{\rm ini} =\prod\nolimits_{n=1}^N\ket {\rm vac}_n$, which is the ground state for $r=0$ if the initial detuning $\Delta_{\mathrm{ini}}$ is positive and large compared to the coupling strength. We then ramp up the scaled driving amplitude $r$ linearly to its maximal value, $r(t)=(t/\tf) r_{\mathrm{max}}$, where $\tf$ is the ramp time. Simultaneously the detuning is linearly decreased to $0$, i.e. $\Delta(t)=\left(1- t/\tf \right) \Delta_{\mathrm{ini}}$. All other
parameters are kept constant. In the numerical plots, all energies and frequencies are in units of $K$.

We are interested in the state of the system at the end of the sweep.
If the oscillators are uncoupled and the sweep is fully adiabatic, the
state of each of them for not too small $r$ will be a symmetric
superposition of states $\ket\psi_j$ ($j=0,1,2$) localized on the
phase plane $(X,Y)$ in Fig.~\ref{fig:cat} at the minima of the
Hamiltonian function $H_0(X,Y)$~\cite{Zhang2017}.  The states
$\ket{\psi}_j$ correspond to classical period-3 oscillations with the
phases that differ by $2\pi/3$ for different $j$.  We associate
$j=0,1$ and $2$ with the directions $0,2\pi/3$ and $4\pi/3$ on the
phase plane toward the wells of $H_0$, respectively, or equivalently,
with the number of the well. If the oscillator is in the state $j$,
the POVM measurement will give the probability
$p_{j'}\approx \delta_{jj'}$.

Coupling the oscillators leads to correlations between their
oscillation phases to minimize the coupling energy. Without the drive
($r=0$), the energy of an individual oscillator is independent of its
phase,  
whereas the multi-oscillator state is invariant only with respect to the continuous global phase, the rotation operator $\exp(-i\theta \sum_n a^\dagger_n a_n)$ commutes with $H_s+H_i$. For $r=0$ and $|V|\ll \Delta_{\rm ini}$ the ground-state multi-oscillator wave
function is the product of the ground-state wave functions of
the individual oscillators, and then $p_{j_1...j_N} = (1/3)^N$.

Not only does the drive break the continuous phase symmetry of an
individual oscillator, but it also reduces the level spacing within
the triples of its neighboring energy levels, see
Fig.~\ref{fig:cat}. Therefore the oscillator coupling becomes
effectively stronger with increasing $r$ and its effect becomes more
pronounced. For large $r$, the low-energy multi-oscillator states are
combinations of the products $\ket\psi_{j_1}...\ket\psi_{j_N}$ of
intrawell states $\ket \psi_{j_n}$ of individual oscillators.  Our
measurement directly reveals such combinations.

\paragraph*{Symmetry arguments.--}

The multi-oscillator initial ($r=0$) state
$\ket\psi_{\rm ini}=\prod\nolimits_{n=1}^N \ket {\mathrm{vac}}_n$
provides the totally symmetric representation of the group generated
by the operators ${\mathbb N}_3$, ${\mathbb T}$, and ${\mathbb R}$.
Since the full Hamiltonian (\ref{eq:full_Hamiltonian}) is invariant
under these symmetry operations, the state $\ket{\psi(t)}$ obtained by
evolving $\ket \psi_{\rm ini}$
will remain totally symmetric. Such a state is {\it not} necessarily
the ground state of the full Hamiltonian. However, it is the {\it
  lowest-energy totally symmetric state}.  If the evolution is slow on
the scale determined by the gaps between the totally symmetric states,
the final state $\ket{\psi(\tf)}$ will be the lowest-energy totally
symmetric state.

\begin{figure}[t]
\includegraphics[width=0.23\textwidth]{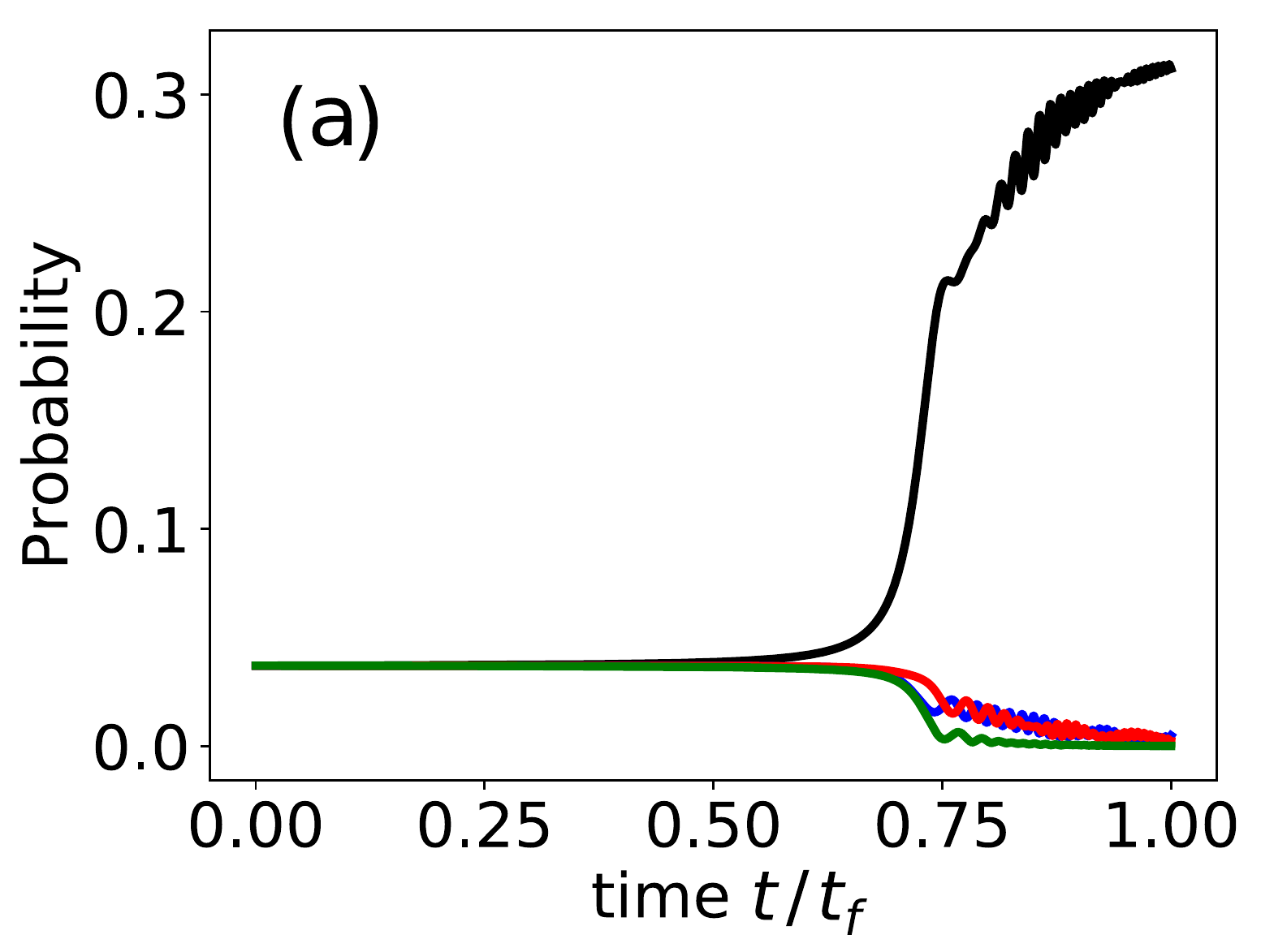} \hfill
\includegraphics[width=0.23\textwidth]{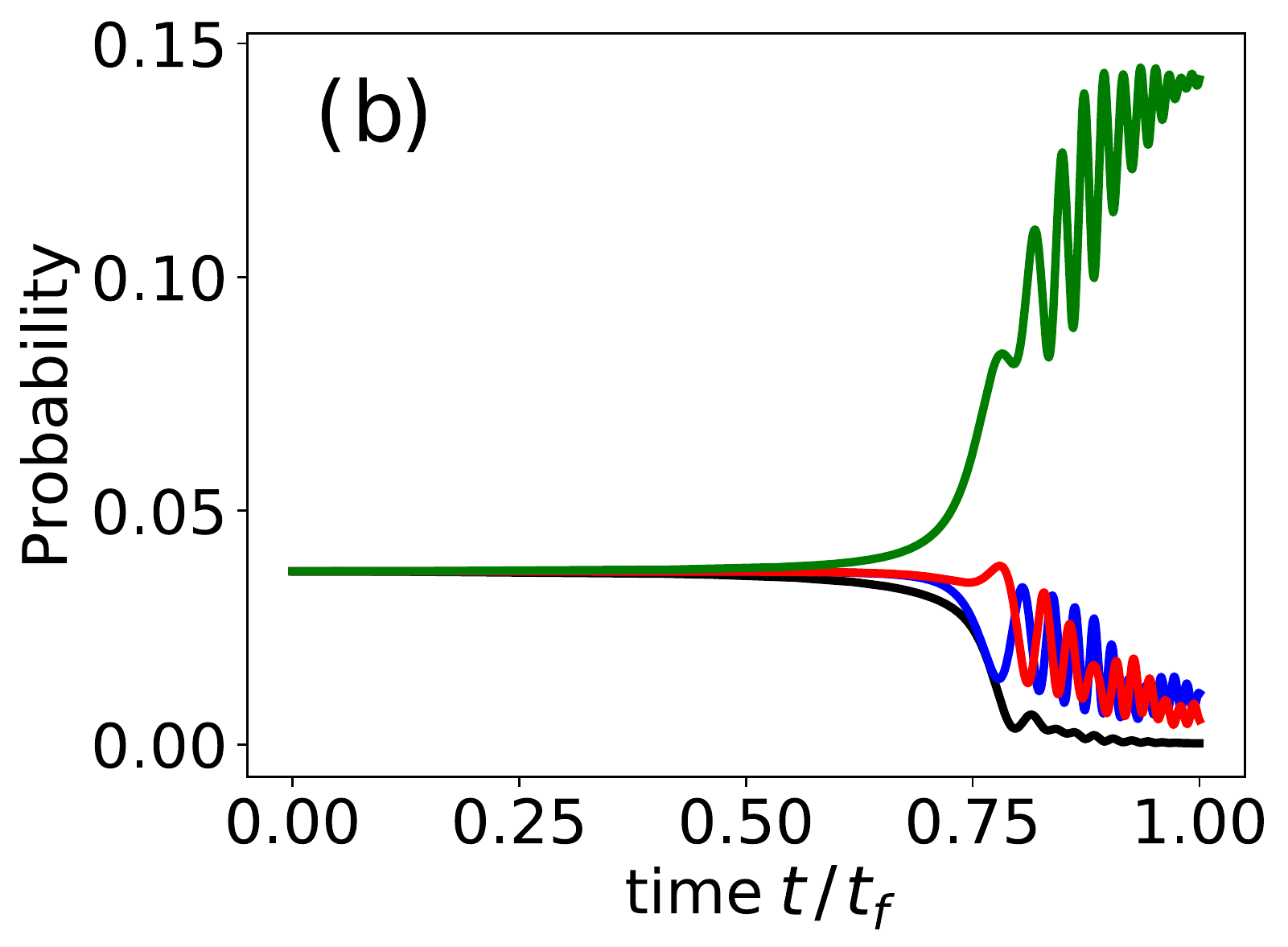}
\includegraphics[width=0.22\textwidth]{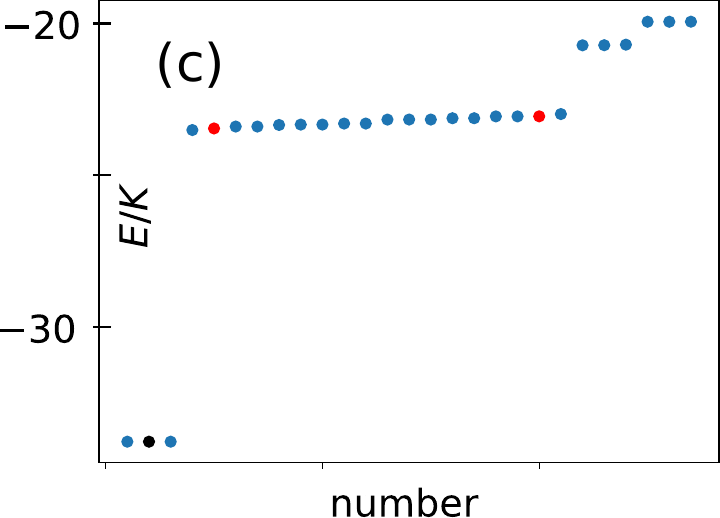} \hfill
\includegraphics[width=0.22\textwidth]{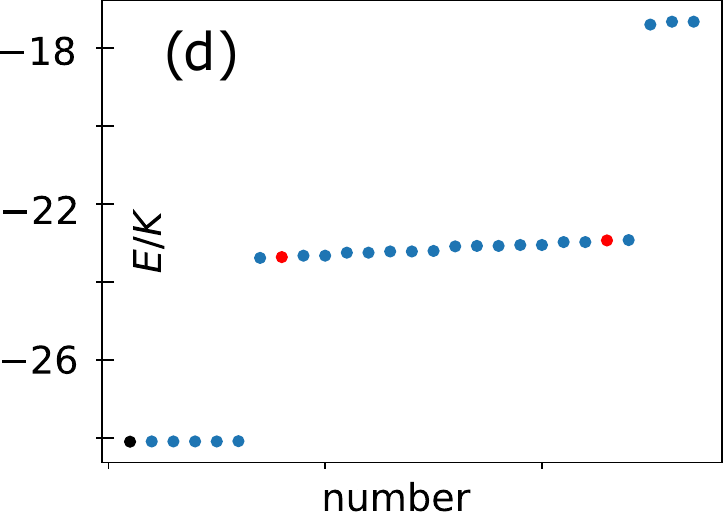}
\caption{Probability evolution and energy spectrum for period tripling
  in a three-oscillator chain with periodic boundary conditions. The
  coupling is ferromagnetic in the left column and anti-ferromagnetic
  in the right column. The parameters are $|V|=0.4K$,
  $\Delta_{\mathrm{ini}}=6K$, the final scaled drive amplitude is
  $r_{\max} = 1.4 K$, and the duration of the sweep is $\tf=100/K$. In
  (a) and (b) the probabilities $p_{jkl}$ of different oscillator
  configurations are encoded as black, blue, red, and green for
  $\{jkl\} = \{000\}, \{001\}, \{002\}$, and $\{012\}$,
  respectively. Due to the geometric phase, the trajectories for the
  configurations $\{001\}$ and $\{002\}$ are different.  Panels (c) and (d) show the 27 lowest eigenvalues of the RWA Hamiltonian (``energies'') at the end of  the sweep. The lowest-energy fully symmetric state is marked black, the next two lowest fully symmetric states are marked red.
}
\label{fig:sweep_tripling}
\end{figure}

\begin{figure}[t]
\includegraphics[width=0.25\textwidth]{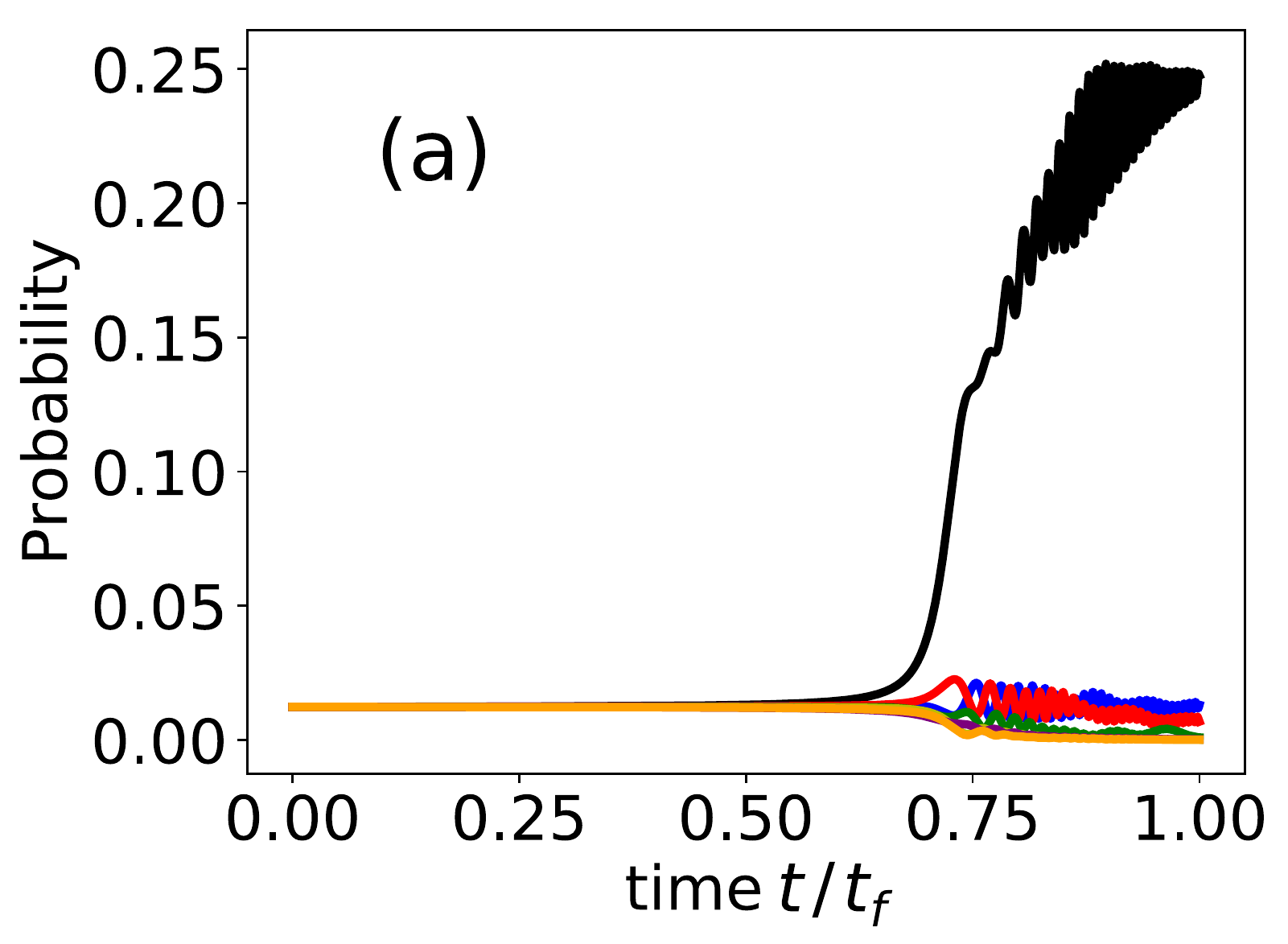}%\hfill
\includegraphics[width=0.25\textwidth]{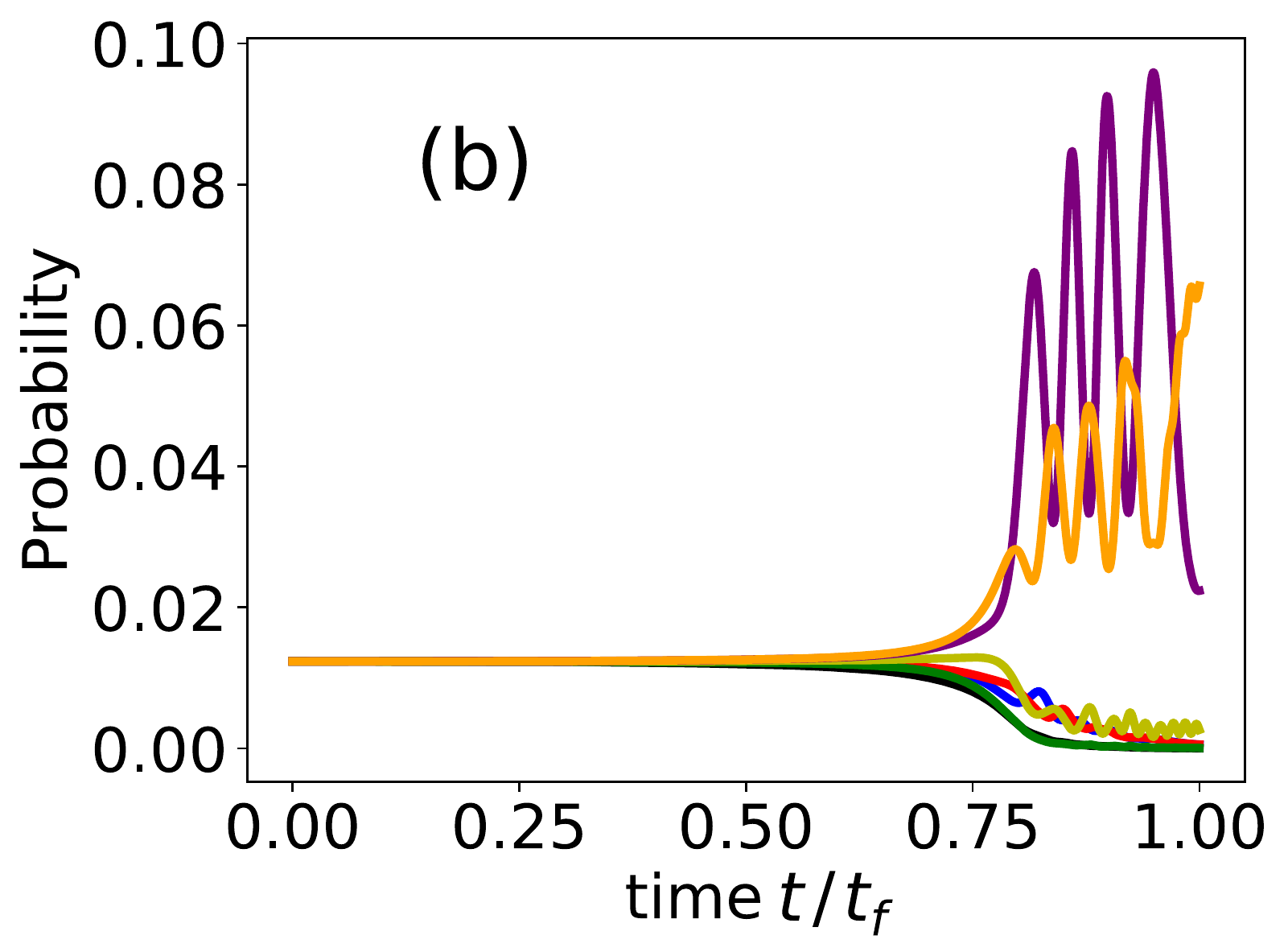}
\caption{Probability evolution for period tripling in a
  four-oscillator chain with periodic boundary conditions. The
  coupling is ferromagnetic in (a) and antiferromagnetic in (b). The
  parameters are the same as in Fig.~\ref{fig:sweep_tripling}. The
  configuration probabilities $p_{jklm}$ are encoded by black, blue,
  red, green, yellow, purple, and orange for
  $\{jklm\} = \{0000\}, \{0001\}, \{0002\}, \{0011\}, \{0012\},
  \{0101\}$, and $\{0102\}$, respectively. The trajectories within the
  initially equivalent pairs $\{0001\}, \{0002\}$, and
  $\{0011\},\{0012\}$ are different. The effect is most pronounced for
  the pair $\{0101\},\{0102\}$ in (b), see main text.}
\label{fig:sweep_tripling2}
\end{figure}

In Figs.~\ref{fig:sweep_tripling} and \ref{fig:sweep_tripling2} we
show, using our POVM-based measurement for a system with three and
four oscillators, that $\ket{\psi(\tf)}$ can be indeed close to the
adiabatic state
 \footnote{The results on the evolution of the symmetric state of two
  oscillators are given in Appendix~\ref{app:two_oscillators}}.

In our simulations the driving parameter $r$ was ramped up to
$r_{\max}=1.4 K$. As seen from Fig.~\ref{fig:cat}~(c), for these
values of $r$ and $\Delta=0$ the three lowest energies of a single
oscillator are close to each other and turn into the tunnel-split
energies of the linear combinations of the intrawell states of $H_0$.

The products of weakly perturbed intrawell states of individual
oscillators $\ket\psi_{j_1}...\ket\psi_{j_N}$ can be denoted as
$\{j_1... j_N\}$, where $j_n$ refers to the $n$th oscillator. To first
order, the coupling energy in such a state is
$ -  V\sum_m (X_{j_m}X_{j_{m+1}}+Y_{j_m}Y_{j_{m+1}})$
where $(X_j,Y_j)$ is the position of the $j$th minimum of $H_0$ on the
phase plane. The operators ${\mathbb N}_3, {\mathbb T}$ and $\mathbb
R$ can be thought of as shift operators in the space of
$\{j_1... j_N\}$, 
\begin{align}
\label{eq:shift_configurations}
&{\mathbb T}\{j_1 j_2... j_N\}=\{j_N\,j_1\,... j_{N-1}\}, \quad {\mathbb R}\{j_1\,j_2... j_N\} = \{j_N\,j_{N-1}... j_1\},\nonumber\\
&{\mathbb N}_3\{j_1... j_N\} = \{j_1-1... j_N-1\}.
\end{align}
The totally symmetric state of the coupled oscillators is found in a
standard way by summing the wave functions obtained by repeatedly
applying the operators $\mathbb T$, $\mathbb N$, and $\mathbb R$ to $\prod_k\ket\psi_{j_k}$.

\paragraph*{Configuration symmetry breaking in the transient regime.--}
For the case of ferromagnetic interaction, the probability to find all oscillators aligned along one direction in the ground state, i.e., to be in the configuration $\{jj... \}$ with $j=0,1,$ or 2 for large $r$, is close to $1/3$, independent of the number of oscillators. This probability is indeed approached in the sweep, as seen from the black
lines in Figs.~\ref{fig:sweep_tripling}~(a) and
\ref{fig:sweep_tripling2}~(a). 

For anti-ferromagnetic interaction the situation is more interesting,
as seen from Figs.~\ref{fig:sweep_tripling}~(b) and
\ref{fig:sweep_tripling2}~(b). For three oscillators the configuration
that minimizes the antiferromagnetic coupling energy for large $r$ is
$\{j_1j_2j_3\}$ with all $j_{1,2,3}$ being different from each
other. There are six such configurations. The totally symmetric state
can be obtained by applying successively the symmetry operators
(\ref{eq:shift_configurations}) to the configuration $\{012\}$.
Respectively, for the adiabatic state preparation, the
probability to find the system in one of the configurations
will be 1/6. This is indeed seen in Fig.~\ref{fig:sweep_tripling}~(b). 

For four oscillators, there are two configurations that both minimize the coupling energy for large $r$, to leading
order in $H_i$ and neglecting tunneling. They are $\{0102\}$ and
$\{0101\}$, and the respective totally symmetric states built out of
them. The only difference between
the configurations $\{0102\}$ and $\{0101\}$ is that, in the first of
them, oscillator 4 is in the well rotated clockwise with respect to
the neighboring oscillators, whereas in the second, this oscillator is
in the well rotated counterclockwise. The equivalence of the
configurations is broken by the geometric phase between the intrawell
states.

The energy splitting between the corresponding totally symmetric
states is small, leading to strong nonadiabaticity with varying $r$
and $\Delta$ and to a similar population of the states. In turn, this
leads to the strong slow oscillations of the configuration populations
in Fig.~\ref{fig:sweep_tripling2}(b). The oscillation period
increases as the energy splitting falls off. Which of the totally
symmetric states has a lower energy depends on the values of $r$ and
$\Delta$, similar to the case of a single
oscillator~\cite{Zhang2017}. There are no reasons to expect that the
symmetric combination of these states has the lowest energy.

The effect of the geometric phase is seen also in
Fig.~\ref{fig:sweep_tripling} (a,b). Here, the transient populations of
the would-be equivalent orientations $\{001\}$ and $\{002\}$ are
different. The probability oscillations are more pronounced for
antiferromagnetic coupling, where nonadiabatic effects are 
stronger.

Further insight into the features of the multi-oscillator states is
provided by their energy spectra. The lowest eigenvalues of the
Hamiltonian of a three-oscillator array for $t=\tf$, are shown in
Figs.~\ref{fig:sweep_tripling}~(c) and (d). Out of 27 states (combinations
of three intrawell states of three oscillators) one can form four
fully symmetric states. Three of them are occupied both for ferro- and
antiferromagnetic coupling. For the fully adiabatic evolution only the
lowest energy one (the black dot) will be occupied. To first order in
the coupling, its energy is shifted down from the energy of
noninteracting oscillators by $3|V|X_0^2$ and $3|V|X_0^2/2$ for the
ferro- and antiferromagnetic coupling respectively; here, $X_0$ is the
distance of the phase-space minima of $H_0$ from the origin.  These
expressions are an overestimate by $\sim 30\%$ for $r=1.4$. The
excited fully symmetric states (the red dots) are also partly
occupied. In leading order, they have the same energy for ferro- and
antiferromagnetic coupling.

As a test, we studied a frustrated triangle of oscillators, where the
first and the third oscillators are coupled antiferromagnetically, but
the second oscillator is coupled ferromagnetically to the other
two. In the absence of the geometric phase, the configurations
$\{000\}, \{011\}$, and $\{022\}$ would be
 expected to have the same
energy.  However, we found that, for the same parameters
$r=1.4 K, \Delta=0$, and $|V|=0.4K$, the symmetrized configuration
$\{000\}$ has the lowest energy.

\paragraph*{Open period-3 system.--} 
The peculiar features of the quantum-coherent dynamics of period-3
oscillations is expected to have a counterpart in the dissipative
dynamics. Some aspects of this dynamics can be revealed by studying
the stationary distribution of a dissipative oscillator in the
ultra-quantum regime. We assume that the dissipation comes from a term
linear in $a, a^\dagger$ that couples the oscillator to a thermal
reservoir. The dissipation-induced change of the density matrix
$\partial_t\rho$ is described by the standard operator
$ \hat L \rho = \frac{1}{2}\kappa(2a \rho a^\dagger - a^\dagger a \rho
-\rho a^\dagger a) $; here $\kappa$ is the energy decay rate, and we
have set the oscillator Planck number $\bar n =0$.

The difference between the classical and quantum dynamics is most
easily seen from the equation for the Wigner distribution 
$W(\alpha,\alpha^*)$. It can be derived in a standard way~\cite{Walls2008}, 
\begin{align} 
\label{eq:Wquantum}
&\partial_t W = 
 \left[
\ i \Delta \partial_\alpha \alpha
  + 2i K \partial_\alpha \alpha ( |\alpha|^2-1)
  - i\frac K 2 \partial_\alpha^2 \partial_{\alpha^*} \alpha \right] W 
\\
&+\left[
  -ir \left(3\partial_{\alpha} (\alpha^*)^2 +
  \frac 14 \partial_{\alpha}^3 \right)
  +\frac{1}{2}\kappa (\partial_\alpha \alpha +
  \frac 12 \partial_\alpha \partial_{\alpha^*}) \right] W \nonumber 
+{\rm c.c.}.
\end{align}
Here, the terms with the first-order derivatives describe classical dynamics in the rotating frame in the absence of quantum fluctuations. For $(3r/2K)^2 > \sqrt{(2-\Delta/K)^2+(\kappa/2K)^2}-2+\Delta/K$, the classical oscillator has three stable states with nonzero $|\alpha|$; they correspond to period-3 oscillations in the lab frame; the state at $\alpha=0$ is also stable.
\begin{figure}[h]
\includegraphics[width=0.23\textwidth]{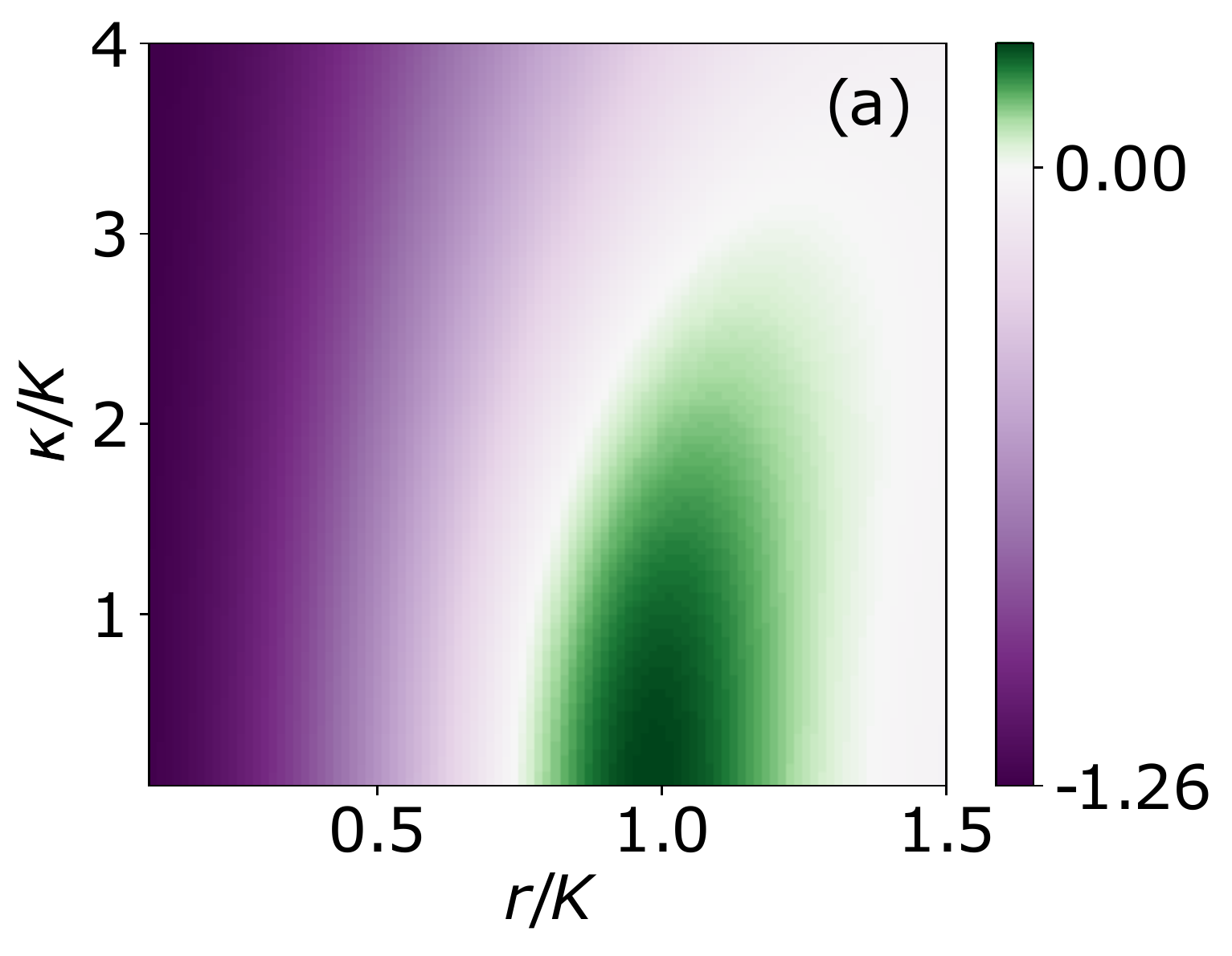}
\includegraphics[width=0.23\textwidth]{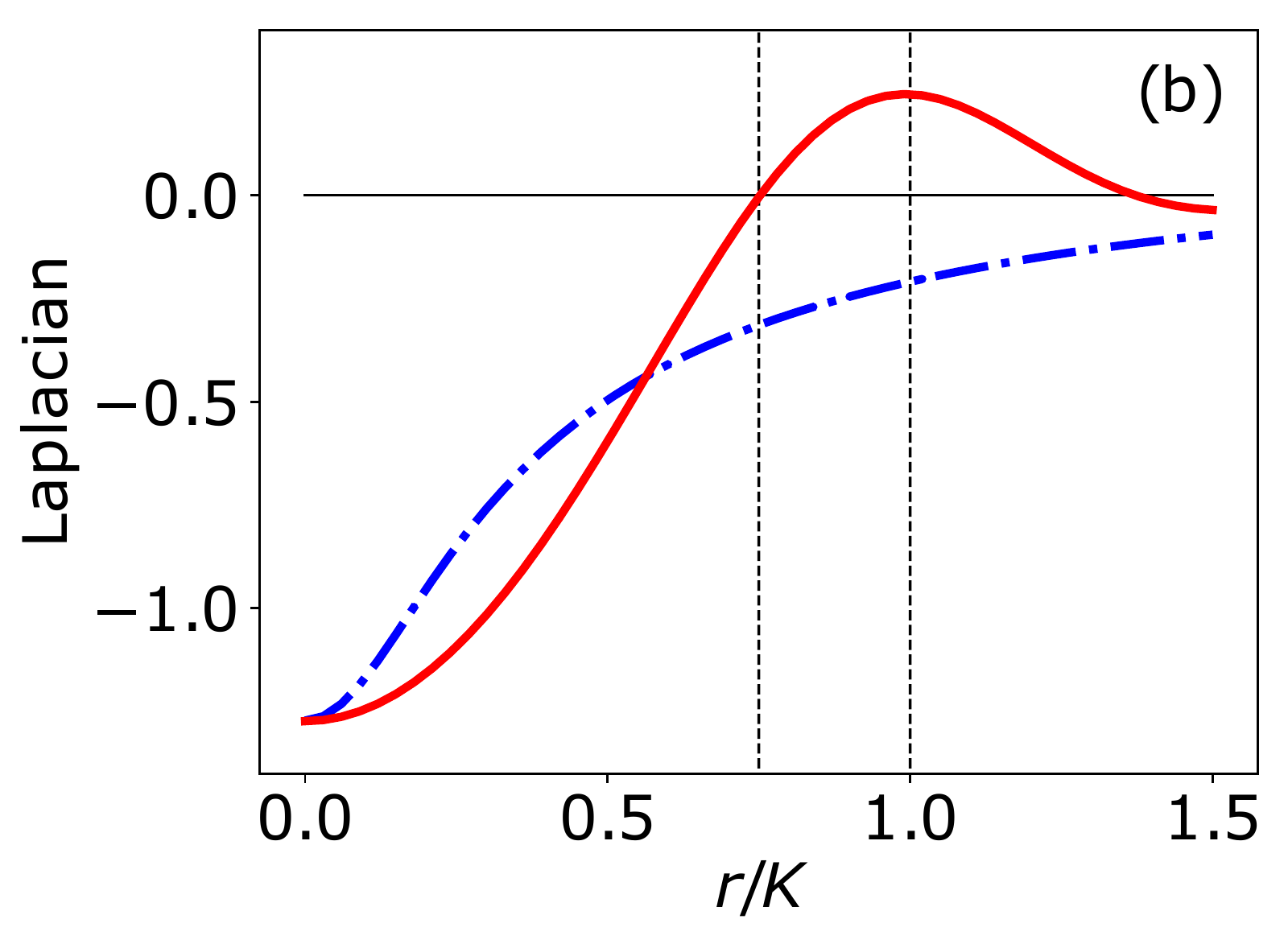}
\includegraphics[width=0.23\textwidth]{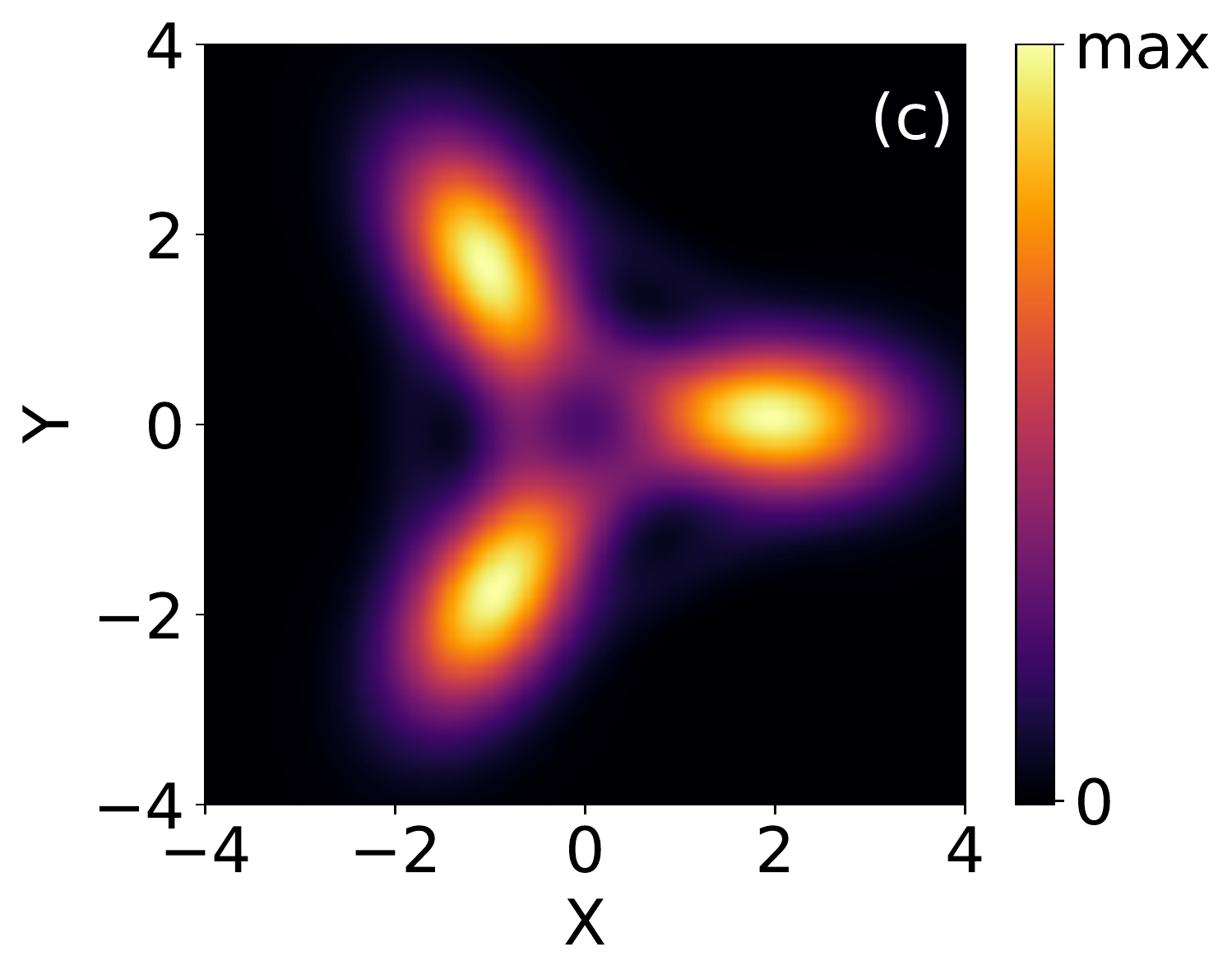}
\includegraphics[width=0.23\textwidth]{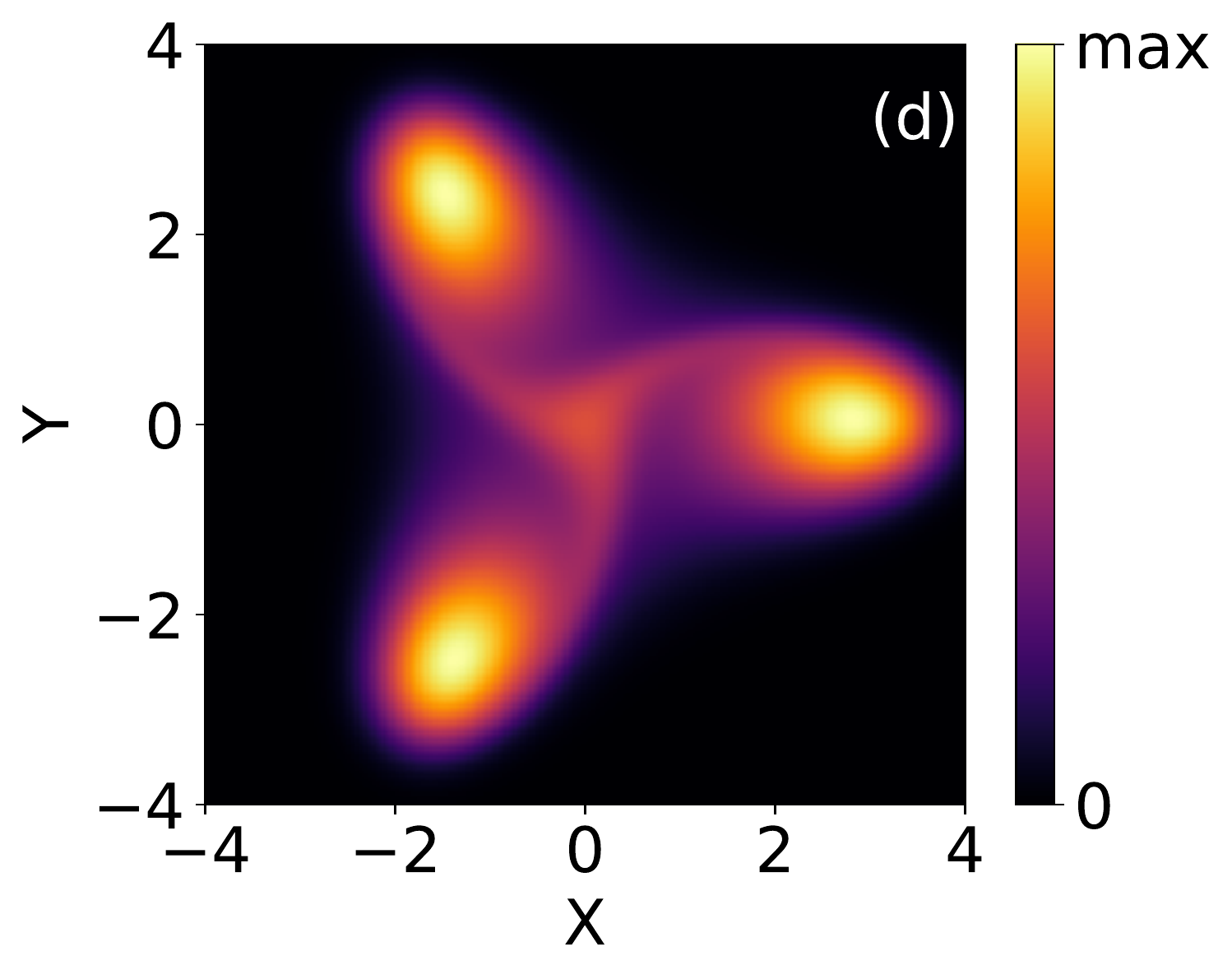}
\caption{
(a) Laplacian of the Wigner density at
the origin of phase space in the full quantum case at zero detuning, as a function of $r$ and $\kappa$ at $\bar n=0$. The green regions
corresponds to positive values of the  Laplacian  which is in contrast
to the classical expectation. (b) Cut at  $\bar n=0$ and $\kappa=0.5K$. The blue dash-dotted line represents the classical baseline, the red full line is the quantum result, where surprisingly a Laplacian $>0$ is possible. The Wigner densities at the dashed black lines for $r=K$, where the difference is most striking, are shown in the lower panels for the quantum case (c) and the semiclassical approximation (d). The Wigner densities at $r = 0.75K$ are given 
in Fig.~\eqref{qTripling} of the Appendix.
\label{fig:wigner}
}
\end{figure}

Within the classical theory one expects the stationary probability
density to display peaks at the stable states. A classical description
refers to the case where $|\alpha|$ varies on the scale $\gg 1$ and
corresponds, in particular, to disregarding the terms with the third
derivatives in Eq.~(\ref{eq:Wquantum}).  However, we found that in the
ultra-quantum regime, where $|\alpha|\sim 1$ in the classical period-3
states, the third derivatives change the distribution qualitatively.
The maximum at $\alpha=0$ can turn into a minimum, see
Fig.~\ref{fig:wigner}~(a) and (b).  The minimum emerges once the drive
becomes sufficiently strong and is most pronounced for $r/K\sim 1$.
As for all parameters in this article, for $\alpha=0$ the eigenvalues
of the Hessian had the same sign, we can use the sign of the Laplacian
$\partial_\alpha\partial_{\alpha^*}W$ to distinguish whether $W$ is
maximal or minimal at $\alpha=0$.

The local minimum of $W$ at the origin disappears for larger frequency
detuning, higher decay rate, or higher temperature, where quantum
effects are less pronounced, 
see Appendix~\ref{app:wigner},
The small curvature $\partial_\alpha\partial_{\alpha^*}W(0)$ for large
$r$ results from the saddle points of $H_0$ approaching
$\alpha=0$. Therefore quantum fluctuations become strong and wash away
the classical stability of the state $\alpha=0$.

\paragraph*{Conclusions.--}

As seen from the above analysis, for period tripling, driven coupled
oscillators exhibit a quantum transition to a correlated state that is
qualitatively different from the classical transition. For
ferromagnetic coupling, with a slowly increasing drive, a quantum
system adiabatically goes into a correlated state of period-3
oscillations. In contrast, a classical system will stay in the
zero-amplitude state. Interestingly, the probabilities of different
seemingly equivalent transient quantum configurations are different,
hinting at an effect of the intrinsic geometric phase of the
oscillators.  These unusual features of quantum oscillator arrays can
be studied with coupled nanomechanical resonators and optical
cavities. A particularly promising platform is provided by coupled
circuit-QED microwave cavities~\cite{Fitzpatrick2016,Ma2019}, as they
combine strong enough nonlinearities and long coherence times. In a
single cavity, period tripling has already been
observed~\cite{Svensson2017a}.

Another unexpected feature of period tripling is that, in the presence
of dissipation, the stationary distributions of the quantum and
classical oscillators are qualitatively different. In a certain
parameter range, the quantum Wigner probability distribution displays
a local minimum, rather than a maximum at the classically stable
zero-amplitude state.

Our results show that period tripling in quantum oscillators allows
studying new many-body phenomena far from thermal equilibrium, which
have no analog in classical systems and in equilibrium quantum
systems.

\paragraph*{Acknowledgments.--} We are grateful to Gil Refael and Mark Rudner for a valuable discussion.
CB and NL acknowledge financial support by the Swiss SNSF and the NCCR Quantum Science and Technology.  MID's research was supported in part by the National Science Foundation (Grant No. DMR-1806473) and the Moore Scholarship from Caltech. Y.Z. was supported by the National Science Foundation (DMR-1609326). All the quantum simulations have been performed using QuTiP~\cite{qutip2}, the semiclassical partial differential equations were solved with numpy \cite{numpy} and scipy \cite{scipy}.

\appendix

\section{A two-oscillator system.--}
\label{app:two_oscillators}
\begin{figure}[h]
\includegraphics[width=0.23\textwidth]{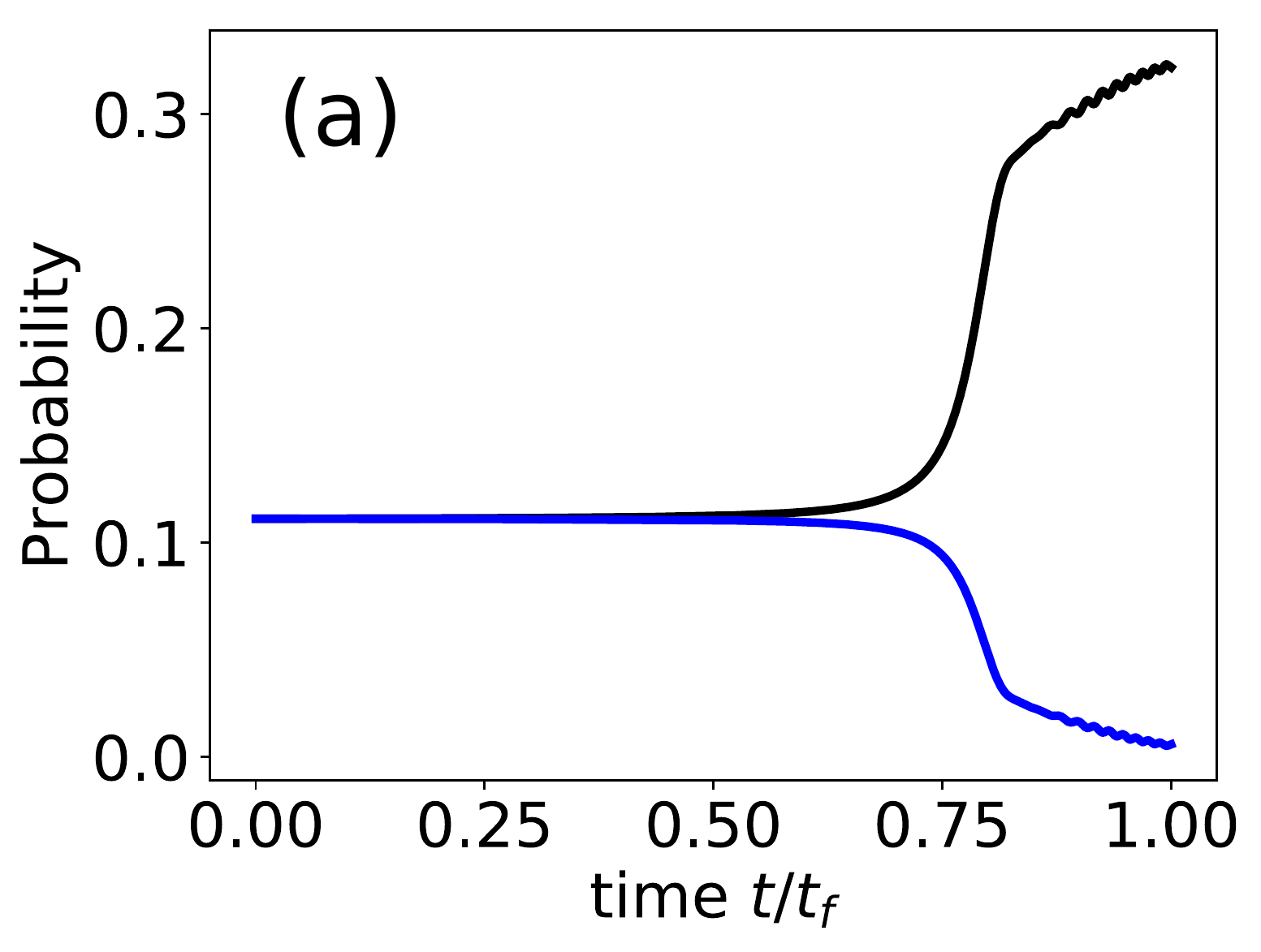} \hfill
\includegraphics[width=0.23\textwidth]{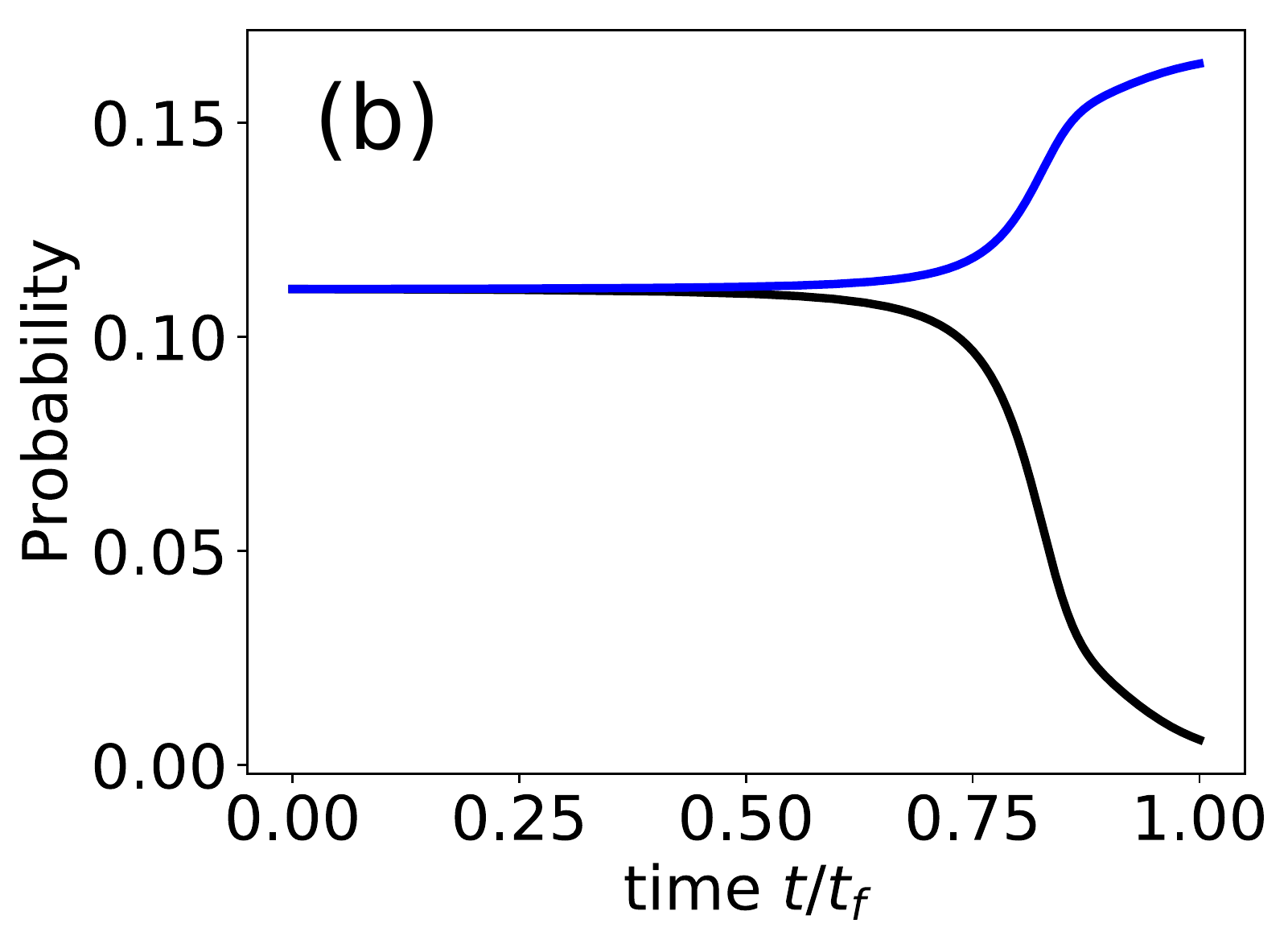}
\includegraphics[width=0.23\textwidth]{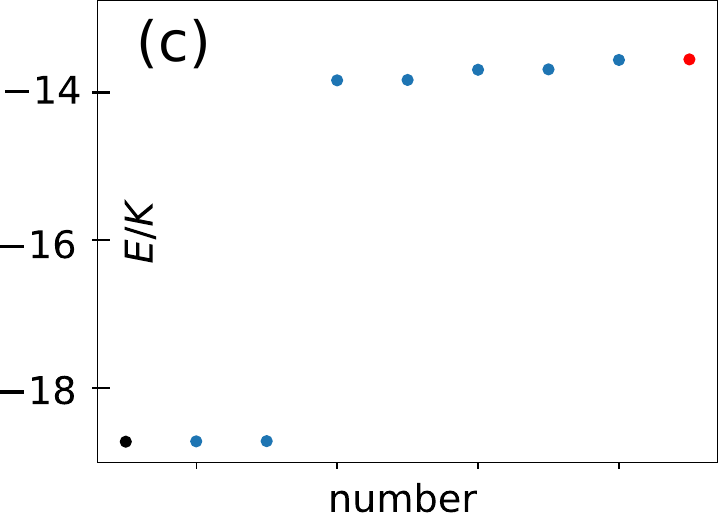} \hfill
\includegraphics[width=0.23\textwidth]{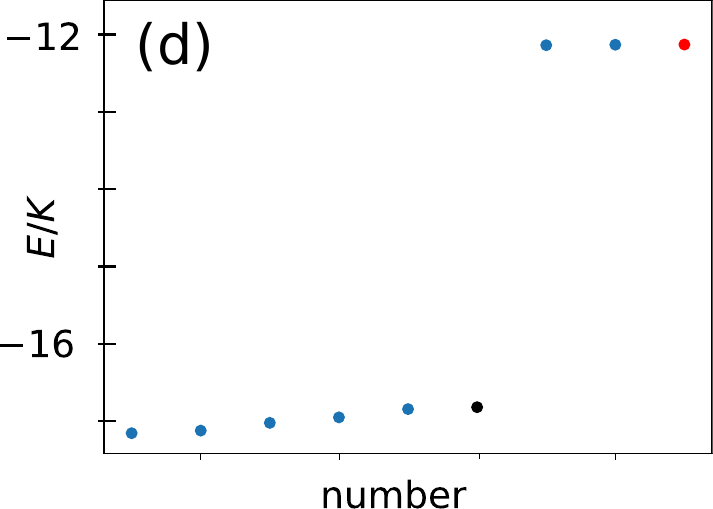}
\caption{ Probability evolution and energy spectrum for period
  tripling in a two-oscillator chain. The coupling is ferromagnetic on
  the left and anti-ferromagnetic on the right. The parameters are
  $|V|=0.4K$, $\Delta_{\mathrm{ini}}=6K$, the final scaled drive
  amplitude $r_{\max} = 1.4 K$, and the duration of the sweep is
  $\tf=100/K$. In (a) and (b) the probabilities $p_{jk}$ of different
  oscillator configurations are encoded as black for \{00\} and blue
  for $\{01\}$. Panels (c) and (d) show the 9 lowest eigenvalues of
  the RWA Hamiltonian at the end of the sweep. The lowest-energy fully
  symmetric state is marked black, the first excited fully symmetric state is marked
  red.}
\label{fig:sweep_tripling_2osc}
\end{figure}

In this section we provide the results that complement the results on period tripling oscillators presented in the main text. Figure~\ref{fig:sweep_tripling_2osc} shows sweeps for two oscillators
that are analogous to the sweeps of three and four oscillators shown
in Figs.~\ref{fig:sweep_tripling} and ~\ref{fig:sweep_tripling2} of
the main article, with the same parameter values. For ferromagnetic coupling, as in the case of 3 and 4 oscillators, the system approaches one of the 3 equivalent configurations $\{jj\}$, which leads to the probability $p_{00}\approx 1/3$. For antiferromagnetic coupling, the most probable configurations are $\{j,j+1\}$, giving $p_{01}\approx 1/6$ at the end of the sweep. 

\section{A strongly non-classical Wigner distribution of a dissipative oscillator.--}
\label{app:wigner}
\begin{figure}[h]
\includegraphics[width=0.23\textwidth]{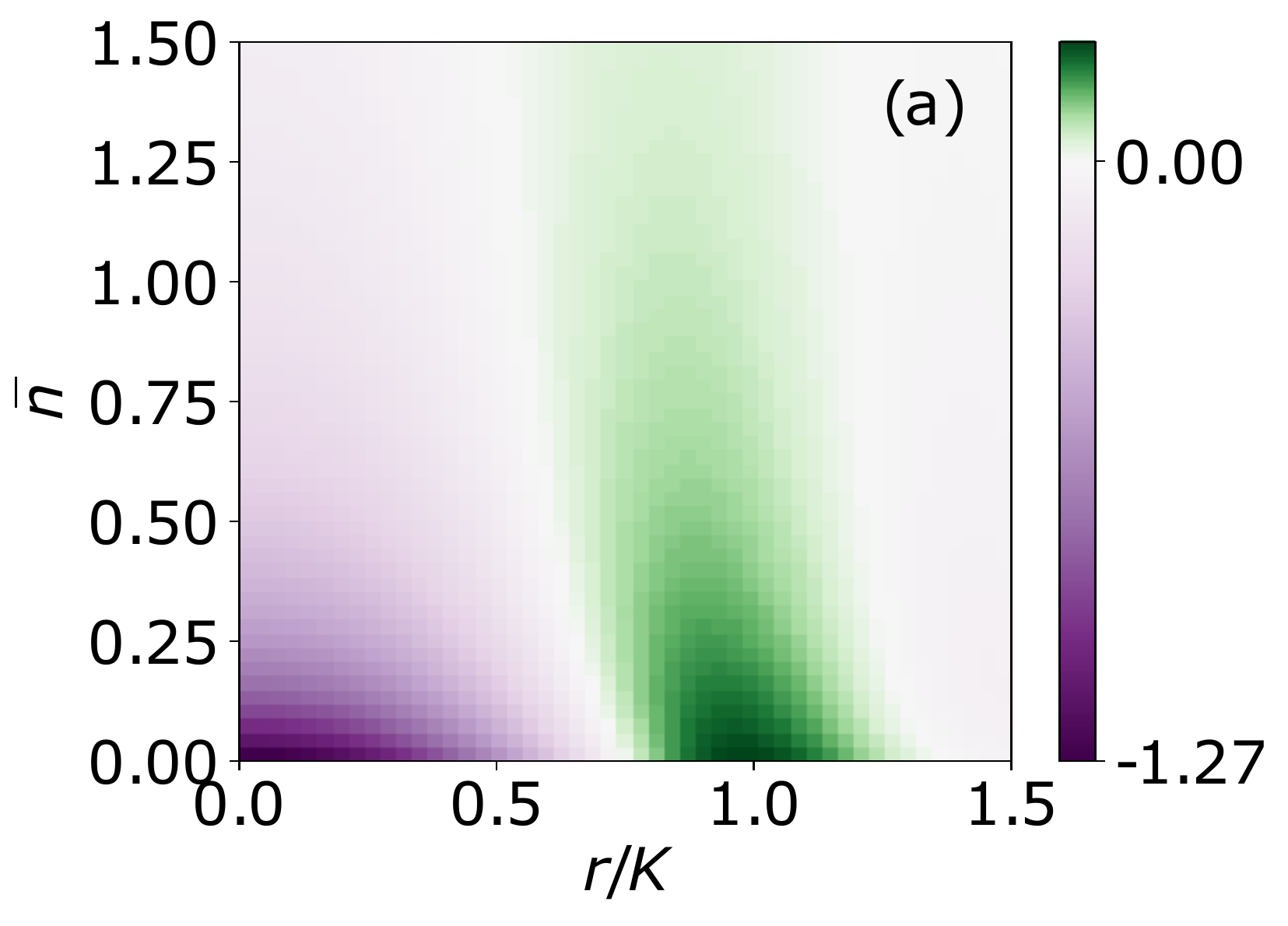}
\includegraphics[width=0.23\textwidth]{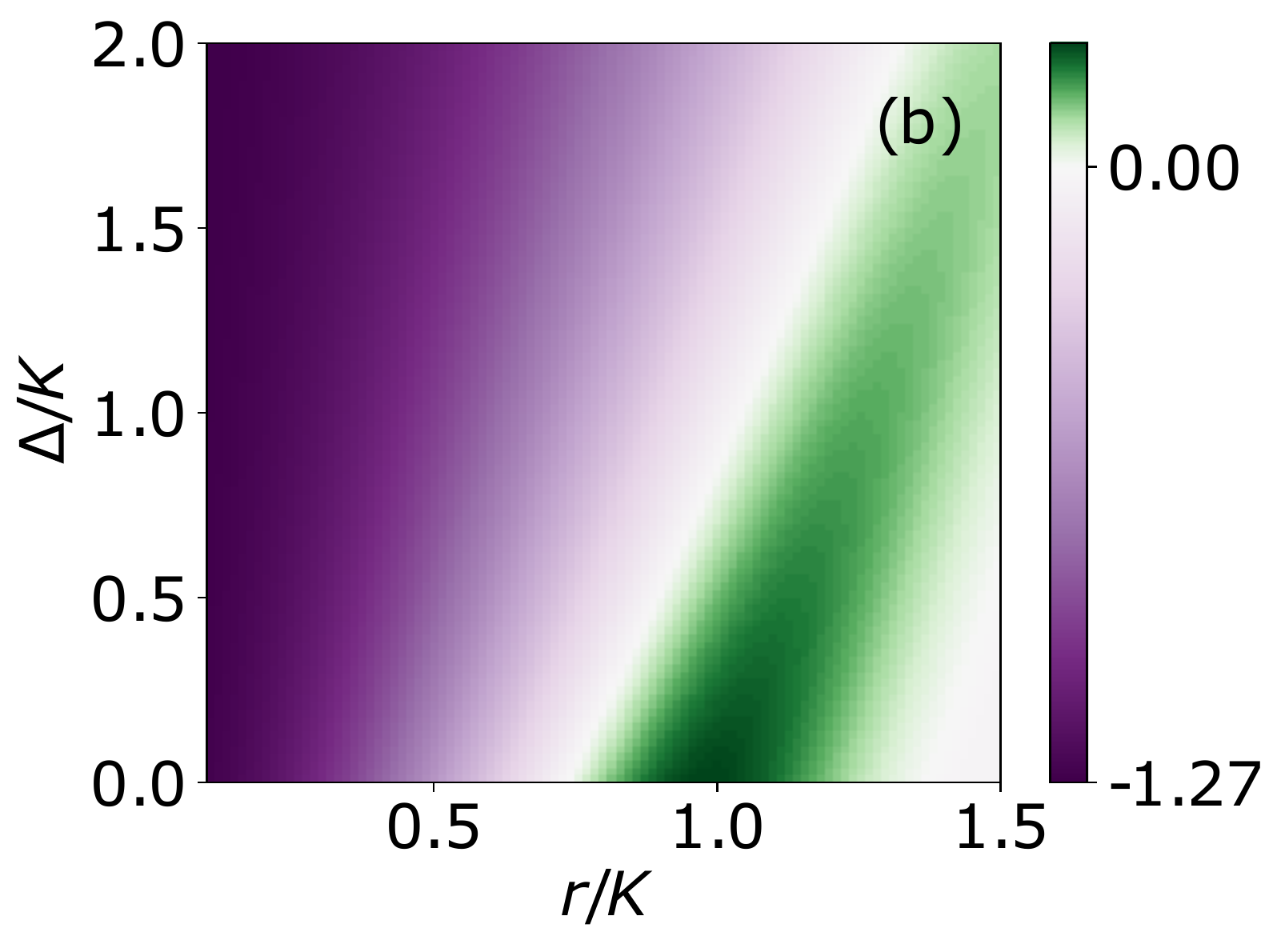}
\caption{
  Laplacian of the Wigner density at the origin of phase space,
  complementing Fig.~\ref{fig:wigner} of the main text.
  Panel (a) is a scan as a function of $r$ and $\bar n$ at
  $\kappa=0.01K$ and $\Delta=0$, panel (b) as a function of
  $\Delta$ at $\bar n=0$ and $\kappa=0.01K$. 
\label{Supp_Scans}}
\end{figure}

In this section we show more detailed results on the region where the
Wigner distribution has a minimum at the classically stable state of
zero vibration amplitude. As explained in the main text, in the
parameter range we have explored, the difference between the maximum
(classical regime) and the minimum (quantum regime) is given by the
sign of the Laplacian of the steady state Wigner distribution at the
origin in the oscillator phase space. Figure~\ref{Supp_Scans} shows
scans of the Laplacian for variable detuning $\Delta$ and variable
Planck number $\bar n$. On increasing $\bar n$, the region of
exhibiting quantum behavior (green area where
$\partial_\alpha\partial_{\alpha^*}W<0$) shrinks, as expected, since
the oscillator becomes more ``classical''. On increasing
the frequency detuning, this area shifts toward larger field amplitudes.

\begin{figure}[h]
\includegraphics[width=0.23\textwidth]{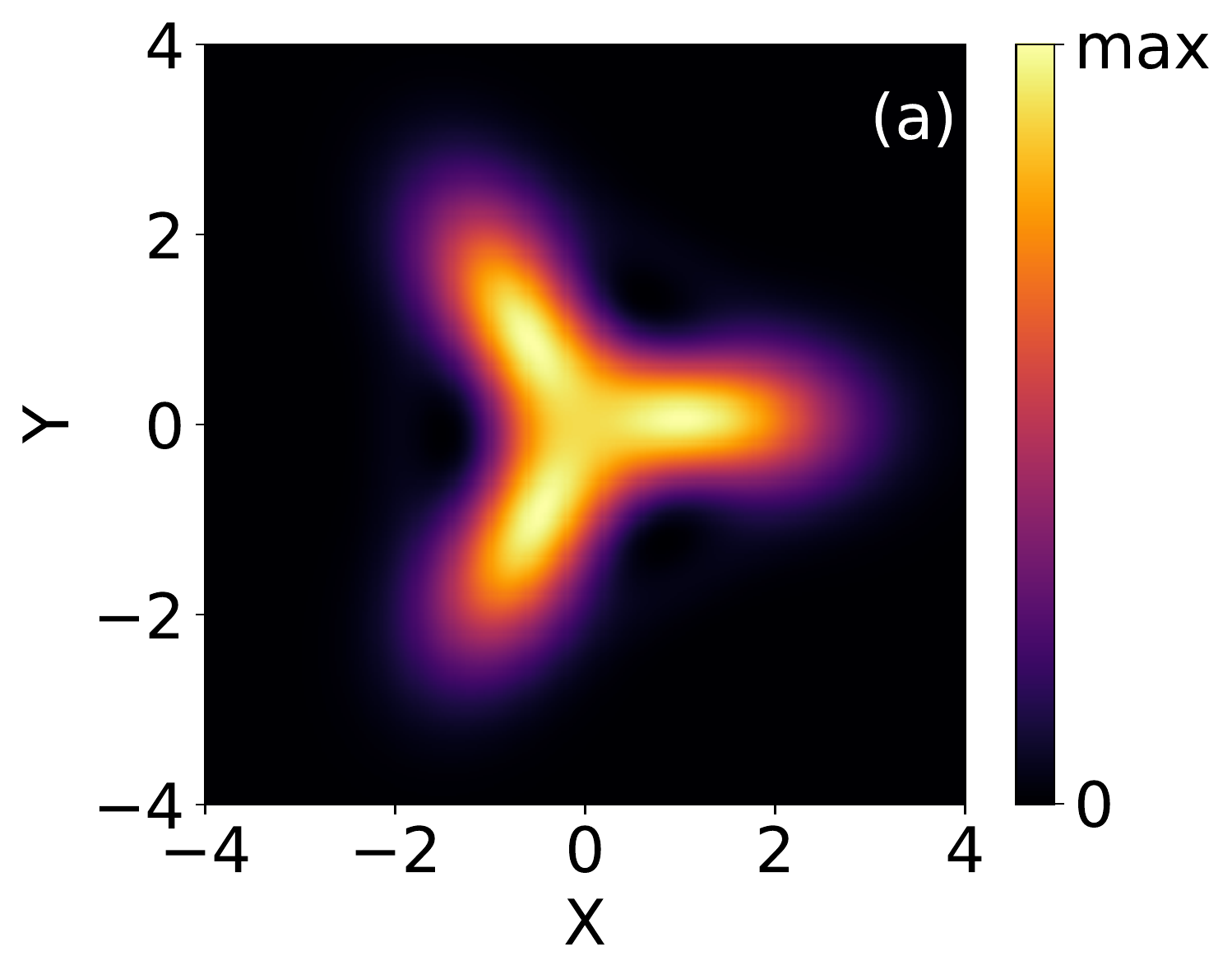}
\includegraphics[width=0.23\textwidth]{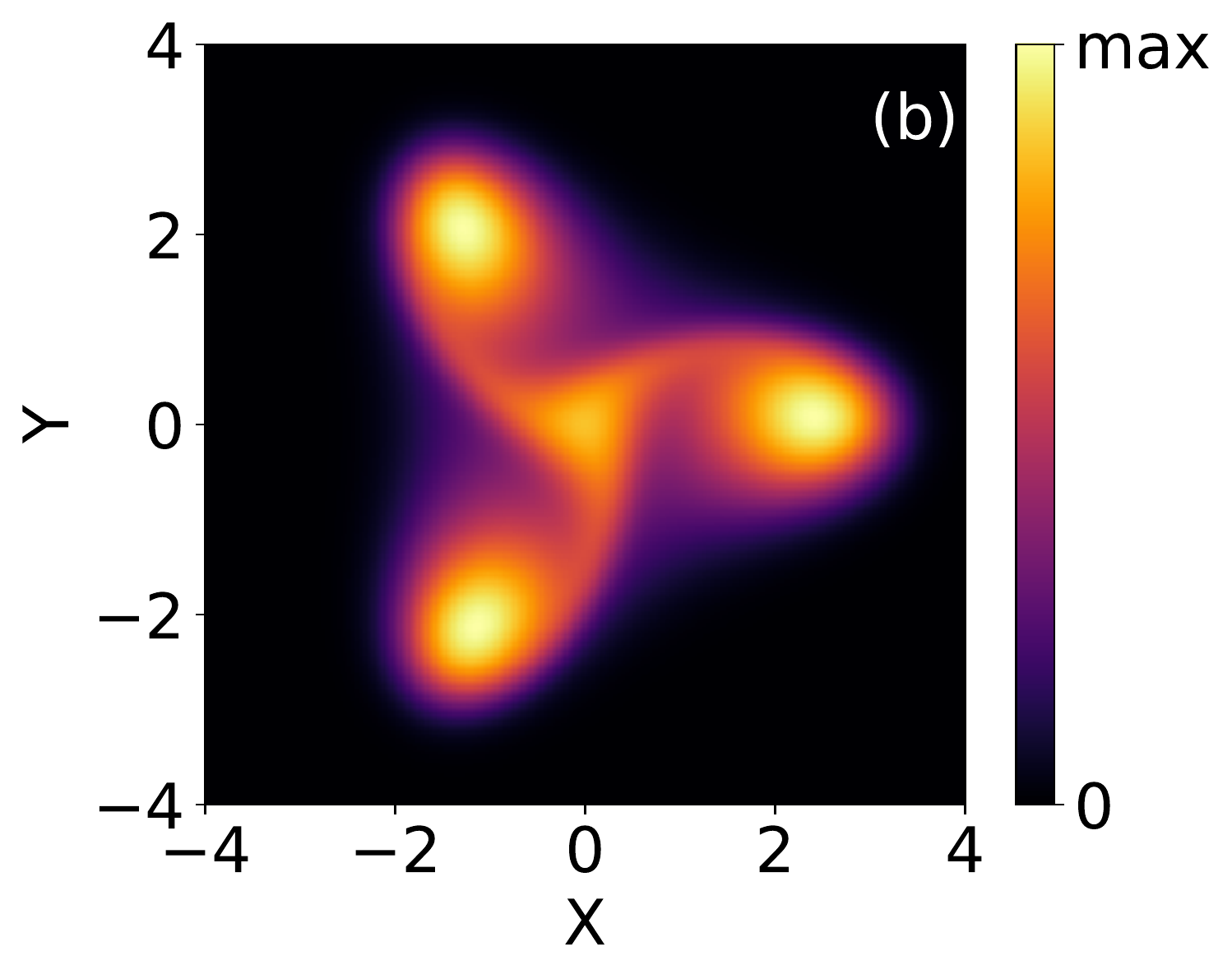}
\caption{
  Steady-state Wigner density of a dissipative oscillator in the
  period-tripling regime for the parameters used in
  Fig.~\ref{fig:wigner}~(b), i.e.,
  $\kappa=0.5K$, $\bar n=0$, $\Delta=0$ and $r=0.75 K$.
  (a) Solution of the full master equation corresponding to
  Eq.~\eqref{eq:Wquantum}.
  (b) Results of the semiclassical approximation
  obtained by solving a Fokker-Planck equation. As can be seen in
  Fig.~\ref{fig:wigner}~(b), the Laplacian in the center is zero for
  the quantum case, while the classical result show three maxima at
  the period-3 states and a maximum at the origin where the oscillator
  amplitude is zero.
\label{qTripling}}
\end{figure}

Figure~\ref{qTripling} illustrates the Wigner density for the
parameters marked at the left dashed line of Fig.~\ref{fig:wigner}~(b).


\section{Comparison to period doubling.--}

For reference, we briefly discuss the case of period doubling, where
the drive Hamiltonian in the rotating frame, given by
Eq.~\eqref{eq:driving} in the main text, is replaced by
\begin{align}
H_d =  -r \sum_n  \left[a_n^2 + (a_n^\dagger)^2 \right]\:,
\end{align}
stemming from a parametric modulation at frequency $\omega_F$ close to
twice the oscillator eigenfrequency $\omega_0$; in this case
$\Delta =\omega_0-\omega_F/2$ in Eq.~(2) of the main text.
 
We map the states of the oscillator to a bit using the measurement
operators introduced in Eq.~\eqref{eq:E_theta} with $P_1= E(\pi/2)$ on the right
half plane and $P_0=\exp(i \pi a^\dagger a) P_1$ on the left
half-plane. The probability $p_j$ for an oscillator in state
$\ket \psi$ to be in bit $j$ is then $p_j=\bra \psi P_j \ket \psi$,
where $p_0+p_1=1$ as expected for the $P$-operators that form a POVM.

\begin{figure}[th]
\includegraphics[width=0.25\textwidth]{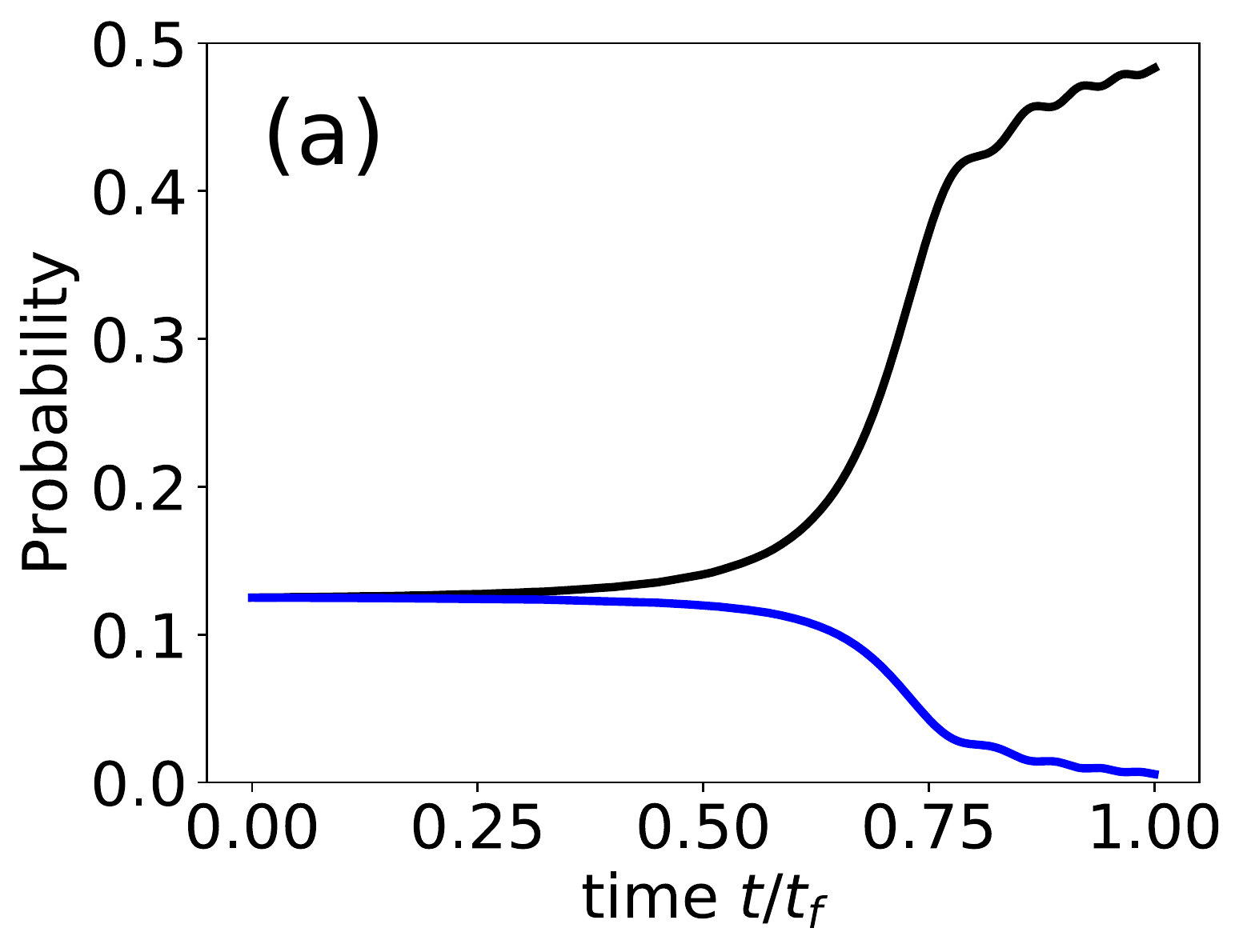}%\hfill
\includegraphics[width=0.25\textwidth]{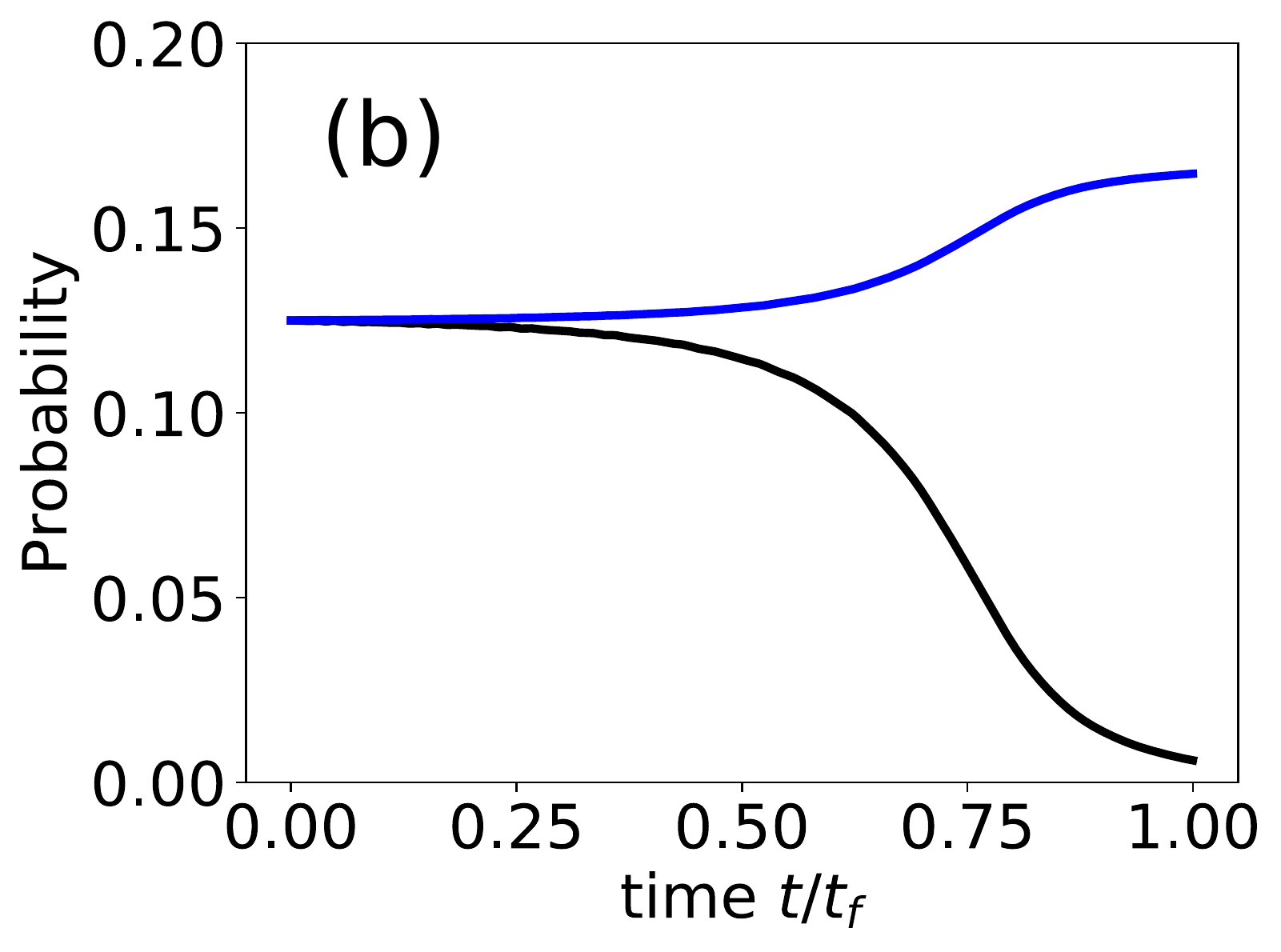}
\includegraphics[width=0.25\textwidth]{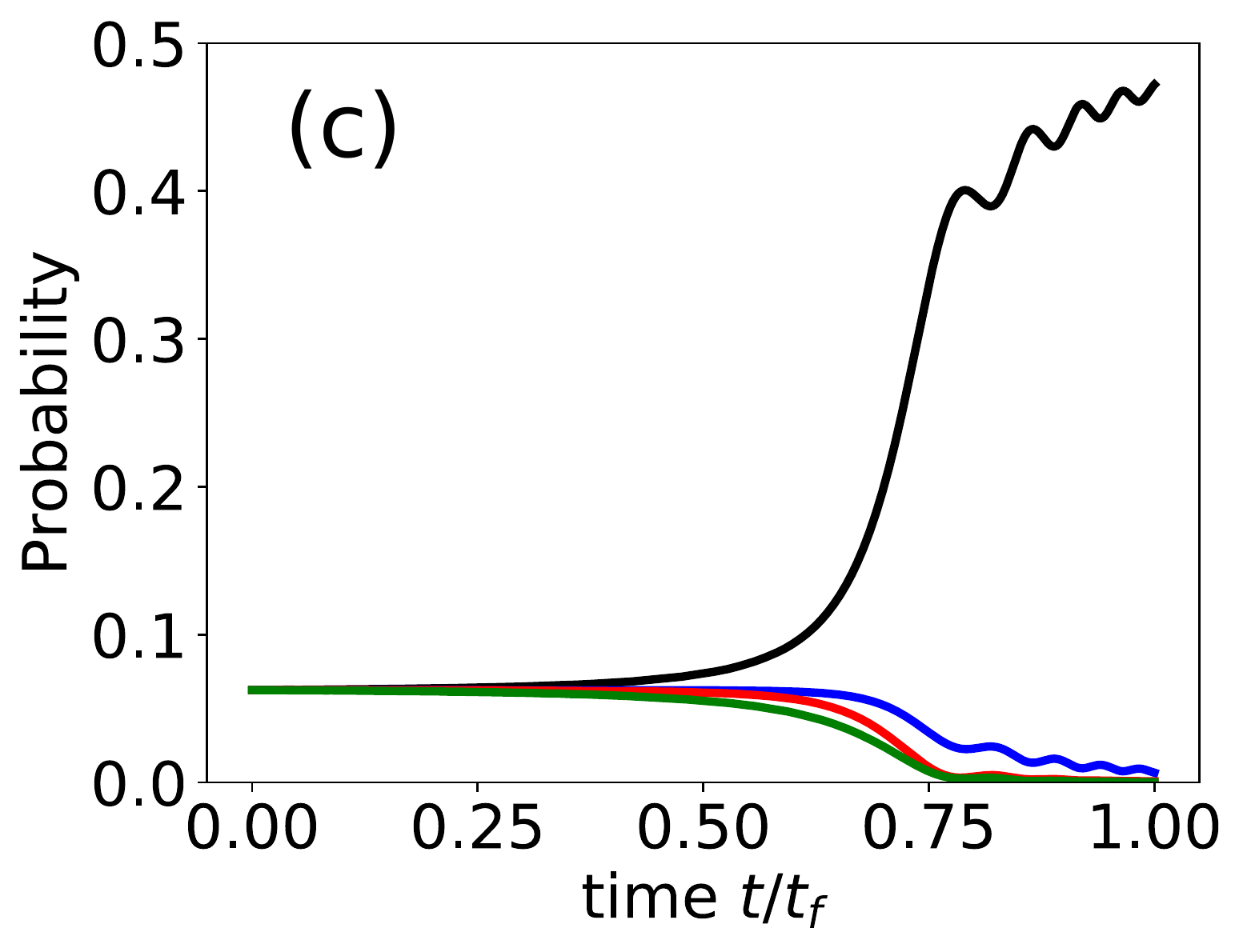}%\hfill
\includegraphics[width=0.25\textwidth]{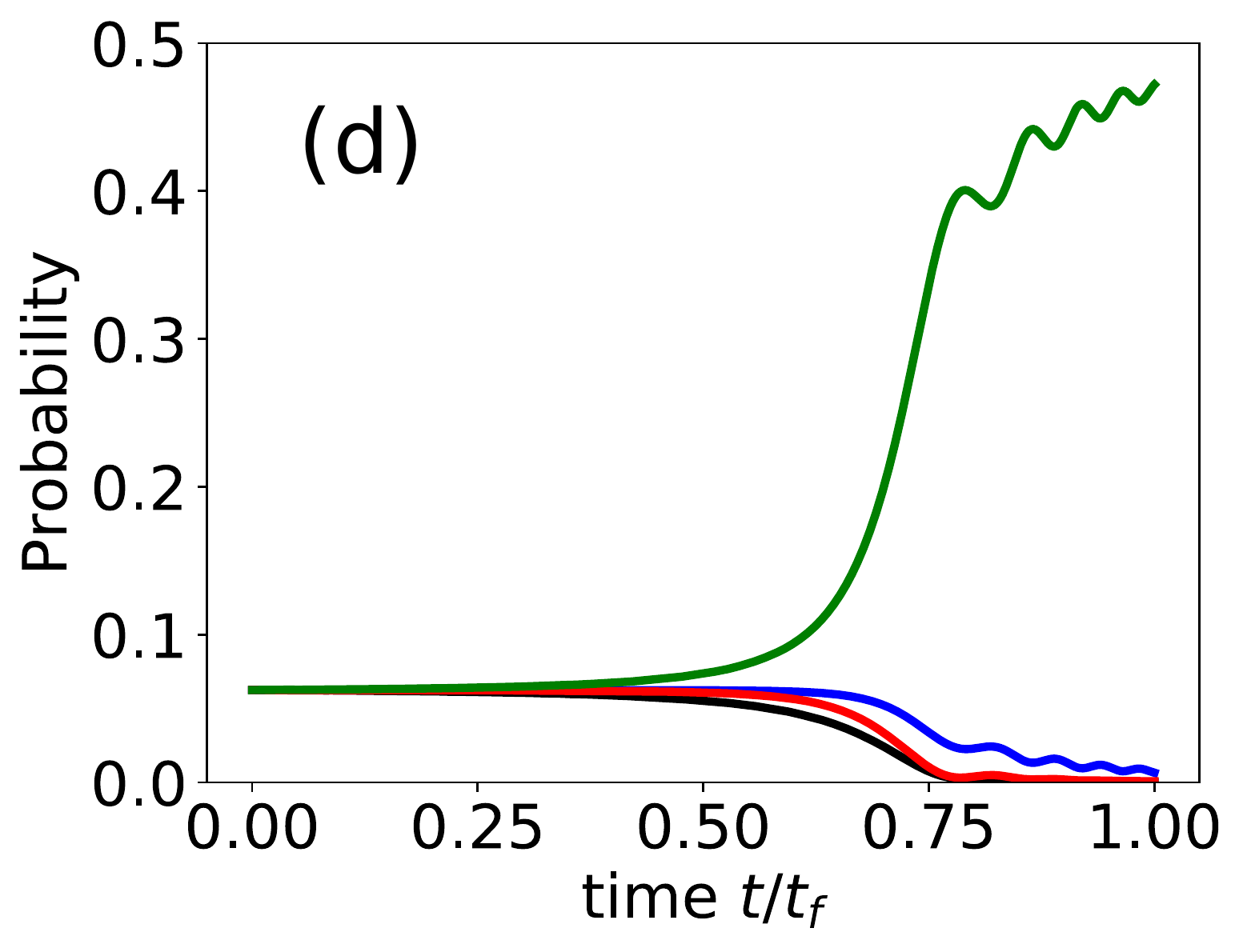}
\caption{Sweep for period doubling for an array of three (upper row)
  and four (lower row) coupled oscillators with periodic boundary
  conditions. The coupling is ferromagnetic in the left column and
  anti-ferromagnetic in the right column. The colors encode the
  probabilities $p_{jkl}$ for $\{jkl\}= \{000\}$ (black) and $\{001\}$ (blue) in
  the upper plots. In the lower plots, the probabilities $p_{jklm}$ of
  the configurations are denoted as
  $\{jklm\}= \{0000\}$ (black), $\{0001\}$ (blue), $\{0011\}$
  (red), and $\{0101\}$ (green).  All other possibilities correspond to one
  of these due to symmetry arguments. The parameters are
  $V=0.4K$, $\Delta_{\mathrm{ini}}=6K$, and $\tf=25/K$ in all plots.
  The maximal driving is $r=2K$.}
\label{fig:sweep_doubling}
\end{figure}

For period doubling the parameters are in a regime close to the
adiabatic limit, so that the maximal probabilities are almost
reached. This maximum occurs at $1/2$, except for anti-ferromagnetic
coupling and three oscillators, where it is at $1/6$. In the
four-oscillator case, the ferromagnetic and anti-ferromagnetic coupling are
equivalent up to a basis transformation, therefore the curves in
panels (c) and (d) of Fig.~\ref{fig:sweep_doubling} agree.

The probability for the system to remain in the lowest fully symmetric
state state or to switch to higher-lying fully symmetric states
crucially depends on the rate of change of the system parameters and
the energy gap to the excited states. In a simplified picture the
system dynamics can be understood as a series of Landau-Zener
transitions occurring at each avoided crossing the system passes
through.  For the Landau-Zener Hamiltonian
$H=\beta^2 t \sigma_z + \Omega \sigma_x$, where $\beta$ parameterizes
the sweep rate and $2 \Omega$ is the minimal energy gap, the
transition probability to the higher-lying state at each of these
crossings is then approximately given by
$P_{\mathrm{LZ}}=1-\exp(- \pi \Omega^2 / \beta^2) $.  In our setup,
increasing $\Delta$ and $V$ leads to larger energy gaps. Together with
the sweep time $\tf$ they fully characterize a sweep. Conveniently, we
can fix $r_{\mathrm{max}}$, as the important transitions only occur
during the phase transition, but not far above threshold, where the
correlations between the oscillators are already effectively
frozen. Also, we already have fixed $K=1$ as a reference, as all other
units are given in units of $K$. It is therefore sufficient to scan
the parameters $(\Delta, V, \tf)$. Due to the numerical cost we focus
on the case of three oscillators.
\begin{figure}[tb]
\includegraphics[width=0.25\textwidth]{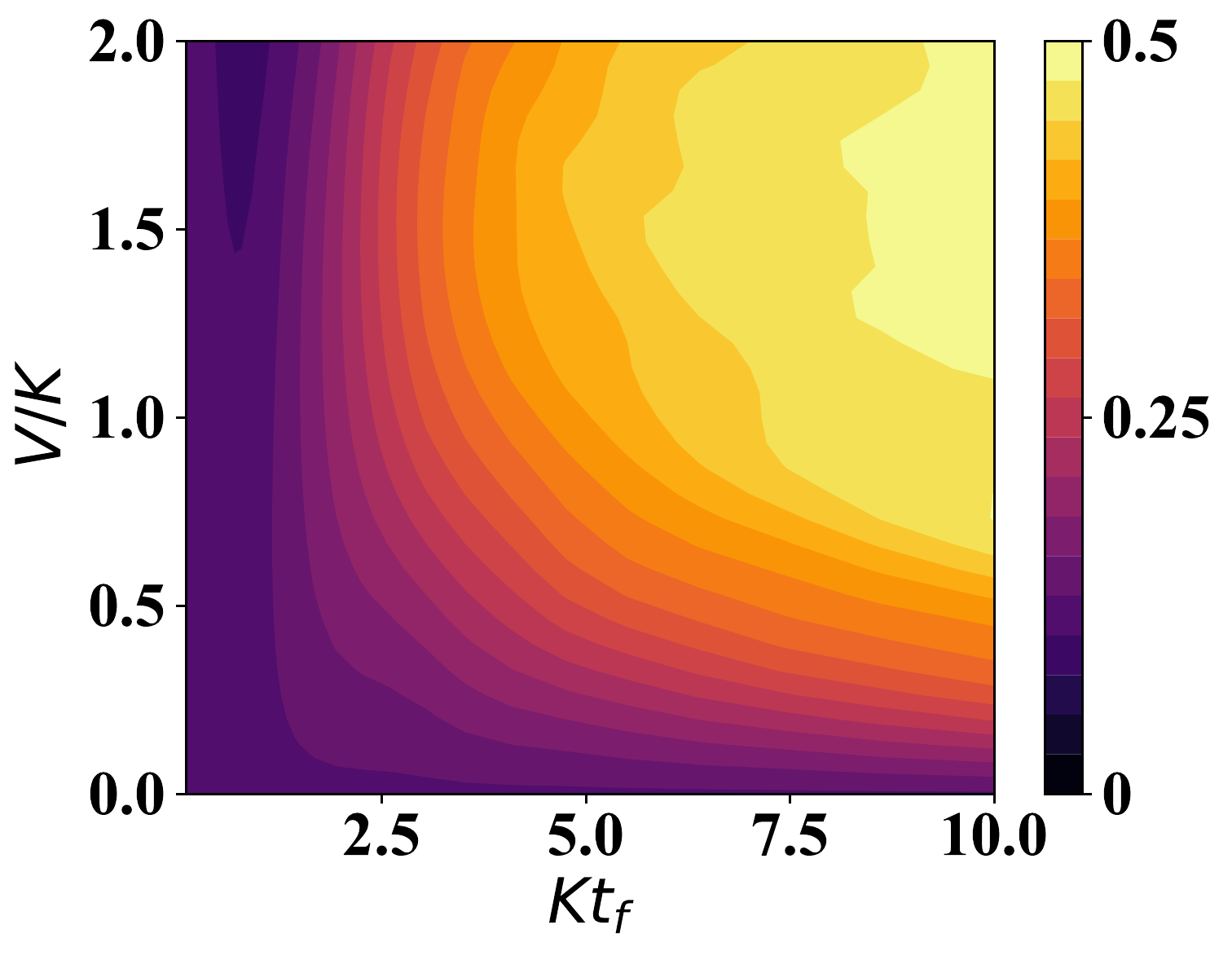}%\hfill
\includegraphics[width=0.25\textwidth]{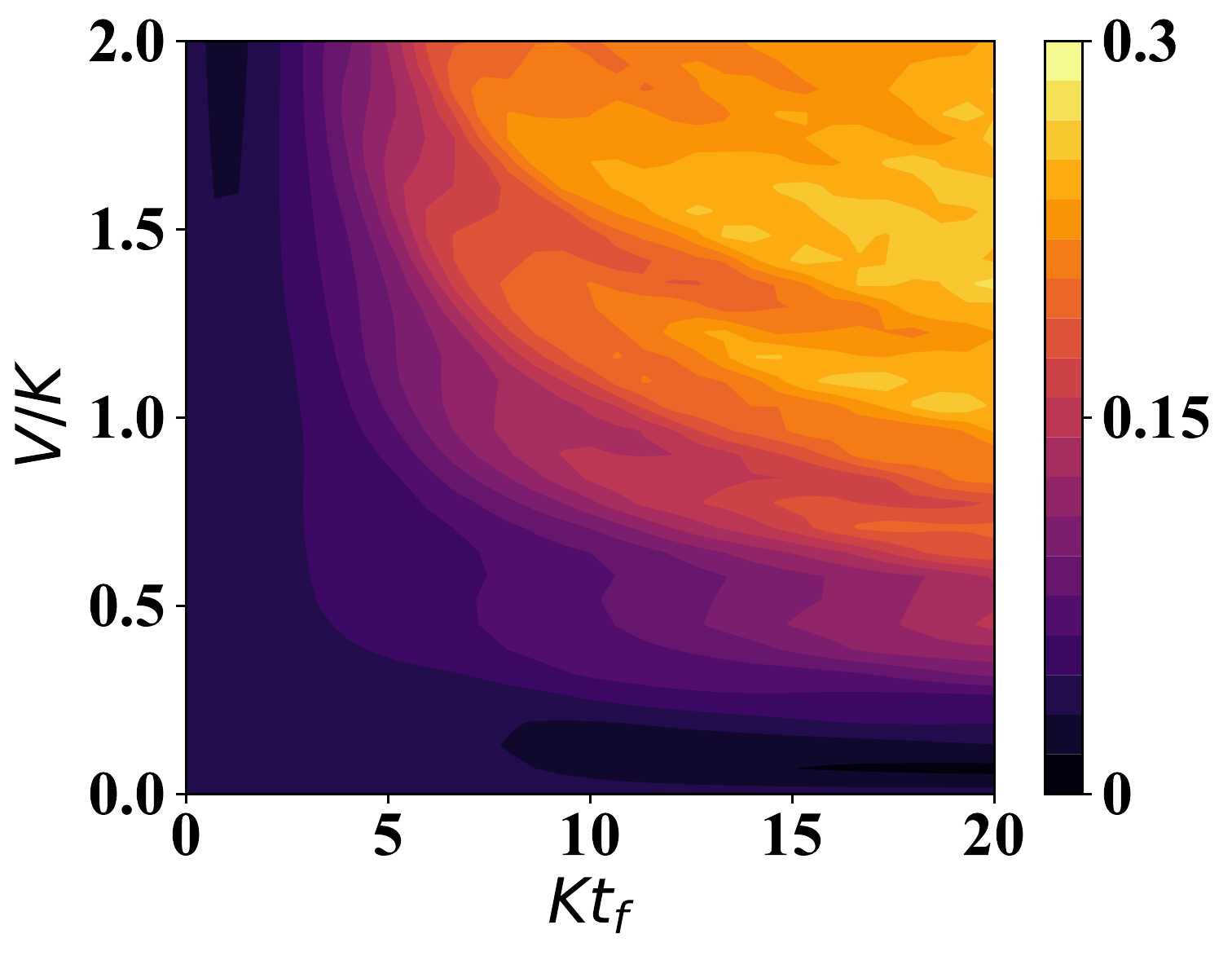}
\includegraphics[width=0.25\textwidth]{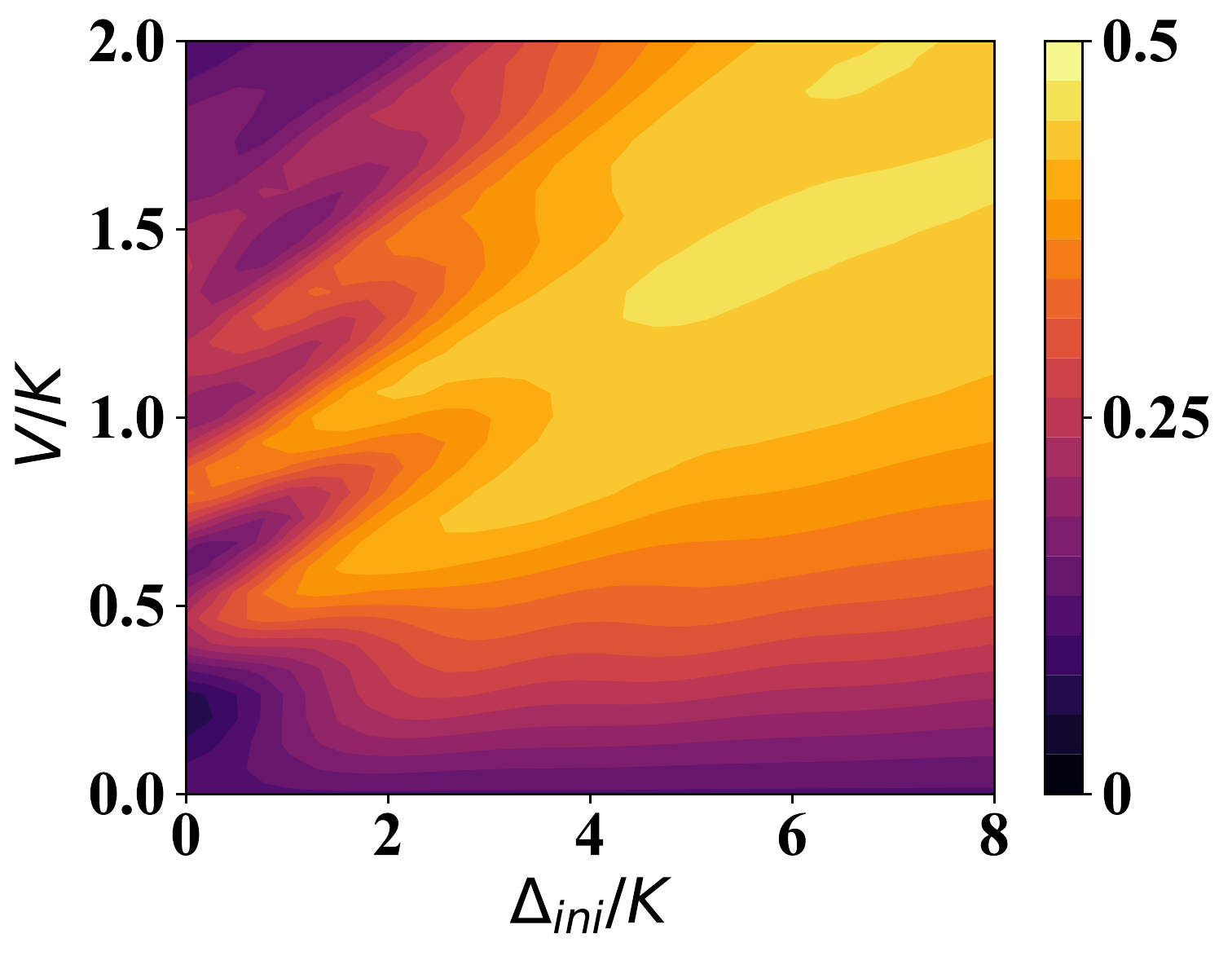}%\hfill
\includegraphics[width=0.25\textwidth]{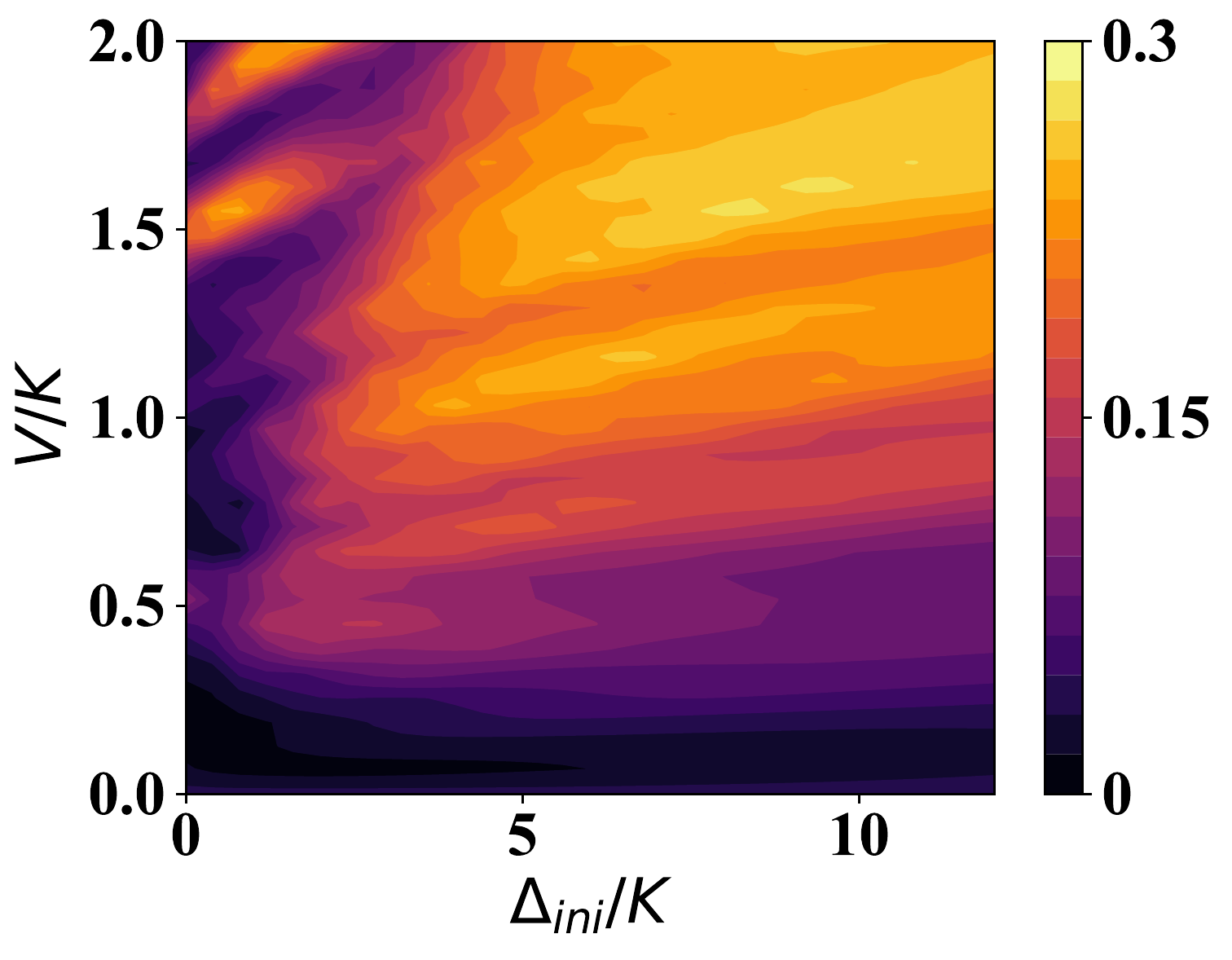}
\caption{Three-oscillator system subject to a full sweep for
  ferromagnetic coupling.  The plots show the probability to end in
  one of the energetically ideal configurations.  Left column: case of
  period doubling for $r=2K$. In the upper plot
  $\Delta_{\mathrm{ini}}=6K$, in the lower plot $\tf=12/K$.  Right
  column: case of period tripling for $r=1.4K$, with
  $\Delta_{\mathrm{ini}}=6K$ in the upper plot and $\tf=30/K$ in the
  lower plot. The oscillators are always initialized in their vacuum
  state, which corresponds to the ground state only for large enough
  $\Delta_{\mathrm{ini}}$. This leads to oscillatory behavior as a
  function of $V$ around $\Delta_{\mathrm{ini}} \approx 0$.}
\label{fig:2dscans}
\end{figure}
As a function of these parameters, Fig.~\ref{fig:2dscans} shows the
probabilities of the configuration minimizing $H_i$ for three
attractively coupled oscillators. For reference, the corresponding
probabilities for period doubling are also shown.  The left panels
refer to period doubling and the right panels refer to period
tripling.  Note that for a perfectly adiabatic sweep, due to,
respectively, the double and triple degeneracy, the maximal
probability to be reached are $1/2$ and $1/3$.

Figures~\ref{fig:2dscans}~(a) and (b) show the threshold of coupling
$V$ and sweep time $\tf$ for the state evolution to remain adiabatic,
panels (b) and (d) show it as a function of $V$ and
$\Delta_{\mathrm{ini}}$. While it is always better to sweep more
slowly, there is a trade-off for the choice of the initial detuning:
The optimal value for $\Delta_{\mathrm{ini}}$ increases for larger
coupling $V$.  For both plots, we conclude that the requirements are
more demanding for period tripling as compared to doubling.

\bibliography{tripling}

\end{document}